\documentclass[a4paper,11pt]{article}
\usepackage{jheppub} % for details on the use of the package, please
\usepackage{subfigure}
\usepackage[T1]{fontenc} % if needed

\title{\boldmath Reduction of the six-dimensional $q$-form fields to the four-dimensional fields by coupling with gravity }

\author[a]{Yong-Tao Lu,}
\author[a,1]{Heng Guo,\note{Corresponding author.}}
\author[a]{Qun Wei,}
\author[a]{and Bing Wei. }
\affiliation[a]{School of Physics,
              Xidian University,
              Xi'an 710071, People's Republic of China}
\emailAdd{luyt@stu.xidian.edu.cn}
\emailAdd{hguo@xidian.edu.cn}
\emailAdd{qunwei@xidian.edu.cn}
\emailAdd{bwei@xidian.edu.cn}

\abstract{
In this paper, we investigate the localization of various $q$-form fields on a codimension-two brane.
In particular, the $0$-form scalar field, the $1$-form $U(1)$ gauge vector field, and
the $2$-form Kalb-Ramond field are considered with gravitational coupling, where a coupling
function $F(R)$ is introduced into the six-dimensional actions of these fields. The
function $F(R)$ depends on the scalar curvature of the bulk. Within this framework,
we find that the massless modes of different $q$-form fields can be localized on the
thick brane for positive values of the coupling parameters $t_1$ and $t_2$.
For the massive modes, the different $q$-form fields exhibit similar localization properties
determined by the coupling parameter $t_2$. When $0<t_2<v^2/24$, the massive
modes of these fields cannot be localized on the brane, while they
may exist as the resonant modes. In the case $t_2=v^2/24$, a finite number of massive
modes can be localized on the thick brane, and the number of localized modes increases with
the coupling parameter $t_1$. Finally, when $t_2>v^2/24$, the effective potentials
associated with the Kaluza-Klein modes of these $q$-form fields form infinitely
deep potential wells, so that all massive modes can be localized on the brane. Moreover, the
tachyonic massive modes can always be excluded for different $q$-form fields.

}

\begin{document}
\maketitle
\flushbottom

\section{Introduction}

The idea that our observed four-dimensional (4D) universe may be a $3$-brane embedded in a higher
dimensional spacetime (the bulk) has provided new perspectives on the gauge hierarchy and cosmological
constant problems \cite{VARubakovplb125136,VARubakovplb125139,MVisserplb15922,KAkamaLNP176267,CWetterichnpb253366,SRDplb16665,JDKjhep06014}.
In the brane-world scenarios, gravity can propagate throughout the bulk, whereas the Standard Model
matter fields are typically confined to the $3$-brane. Consequently, it is natural to consider the
compactification of extra dimensions into a small spatial volume. One of the earliest and
most influential frameworks of this type is the Kaluza-Klein (KK) theory \cite{TKaluza1921,OKlein37895}. Subsequently, Rubakov
and Shaposhnikov proposed the possibility of noncompact extra dimensions through the domain-wall
model \cite{VARubakovplb125136,KAkamaLNP176267}. Later, Randall and Sundrum (RS) introduced the RS
model where the Newtonian potential can be recovered on the brane even though the extra dimension is
infinite \cite{RS834690}.

In early extra-dimensional models, the $3$-brane was usually treated as an infinitely thin object.
However, fundamental theories are generally expected to involve a minimal length scale, implying
that a realistic brane should possess a finite thickness. Motivated by this consideration, thick
brane models with nonvanishing thickness and internal structure were developed by combining features
of the RS model and domain-wall configurations \cite{Csakinpb581309,Gremm478434,DeWolfeprd62046008,Shiromizuprd62024012}.
Since then, thick brane modes with various geometric and field configurations have been extensively
investigated \cite{MGprd62044017,AKmpla171767,AKplb50438,SKprd65064014,AWprd66024024,TRSjhep04062,VDprd77044006,
VDgrg41131,DBjhep11070,DBjhep05012,DBjhep11064,VIAplb634526,LYXjhep10069,AHAjhep11015,LYXjhep06135,
DBplb726523,ZYepjc76321,GHprd107104017,GHplb868139718}.

In brane-world scenarios, gravity is free to propagate throughout the bulk, whereas the matter
fields of our observed 4D universe are confined to the $3$-brane, in agreement with current
experimental observations \cite{VARubakovplb125136,VARubakovplb125139,MVisserplb15922}. To recover
the Standard Model on the brane, the zero modes of various bulk matter fields and their interactions
must be localized on the brane through appropriate mechanisms. In general, massless scalar fields
can be naturally localized on branes of different types, including those in six-dimensional (6D)
spacetime \cite{WJJjhep2105017}. For flat brane configurations, the massless scalar mode can
generally be localized on the brane, whereas the massive modes are typically nonlocalizable.
However, by introducing additional couplings, the massive KK modes may exhibit different
localization behaviors. In refs. \cite{HGuo2310.01451,YTLu2401.11688}, the coupling between
the scalar field and the gravity is considered in five-dimensional (5D) brane models. Under
such coupling, the effective potential for scalar KK modes can asymptotically approach zero,
a positive constant, or positive infinity when far away from the brane. As a result, the
massive modes can be quasi-localized or localized on the brane. The quasi-localized ones
are also referred to as resonances. Recently, the researches on these specific states are
reported with the discussion for their evolution, where an analysis of the feasibility of
the scalar resonances as a dark matter candidate is presented \cite{TQepjc2023}. These works
originally explored the evolution of KK resonances of various fields in thick brane.

For the $U(1)$ gauge vector field, localization can be realized on the RS brane in certain
higher-dimensional models \cite{WJJjhep2105017,IOdaPLB29600113}, as well as on thick de Sitter
branes \cite{HGuo1103.2430} and Weyl thick branes. However, in 5D flat brane case, the
localization of the $U(1)$ gauge field generally requires additional coupling mechanisms. For
instance, the $U(1)$ gauge field is assumed to be coupled to the background scalar field in
refs. \cite{CAVaqyera1406.2892,XNZhang2405.16324} and to the dilaton in ref. \cite{WCruz1211.7355}.
In addition, the geometric coupling mechanism was proposed in refs. \cite{LFFreita1809.07197,GAlencar1409.4396,ZHZhao1406.3098,ZHZhao2212.00444},
which enables the localization of the massless mode of the $U(1)$ gauge field. This mechanism
has also been extended to the study of massive modes which the resonant modes and $p$-form
fields are studied \cite{RRLandim1105.5573,ICJardim1410.6756}. Motivated by these developments,
the authors in refs. \cite{GHprd107104017,YTLu2401.11688,XNZhang2405.16324,ZHZhao1712.09843}
introduced a curvature-dependent coupling function $F(R)$ into the 5D action to describe the
interaction between the $U(1)$ gauge field and the gravity. Within this framework, the massless
mode of the $U(1)$ gauge field can be localized on the brane, while the massive modes may become
either quasi-localized or fully localized on both flat and de Sitter thick branes.

It is well known that the $0$-form and $1$-form fields correspond to the scalar field and the
$U(1)$ gauge field, respectively, while the $2$-form field corresponds to the Kalb-Ramond (KR)
field. In 4D spacetime, the KR field is dual to a scalar field, whereas in higher-dimensional
spacetimes it represents an independent degree of freedom and may describe additional particles.
Under minimal coupling, the KR field cannot be localized on thick flat branes in either the 5D
or 6D spacetimes \cite{LYT2510.16491,YTLu2401.11688}. To overcome this difficulty, various
localization mechanisms based on nonminimal couplings have been proposed. In particular, when
coupled to gravity, the KR field can be localized on a flat brane in 5D spacetime, and its
massive KK modes can appear as resonant states \cite{YTLu2401.11688,XNZhang2405.16324}.
Alternatively, couplings between the KR field and the background scalar field have been investigated
in refs. \cite{XNZhang2405.16324,YZDu1301.3204}, where the localization of the KR field on the
brane was successfully achieved. Such coupling mechanism has also been extended to 6D spacetimes
\cite{LYT2510.16491}. In that case, the KR field can be localized on a flat brane when the
coupling parameter satisfies $t>v^2/12$, and resonant modes may also arise.

In this paper, we investigate the localization of different $q$-form fields, namely the scalar
field, the $U(1)$ gauge vector field, and the KR field, by introducing a coupling between the
$q$-form fields and the gravity in a 6D spacetime. Specifically, a coupling function $F(R)$, depending
on the scalar curvature of the bulk spacetime, is incorporated into the actions of the 6D $q$-form
fields. The two extra dimensions of the bulk consist of one noncompact (large) extra dimension
and one compact extra dimension. Within this framework, we demonstrate that different $q$-form
fields exhibit similar localization behaviors. As both coupling parameters $t_1$ and $t_2$ are
positive, the massless modes of the various $q$-form fields can be localized on the thick flat
brane. For the massive modes, the localization properties are mainly governed by the coupling
parameter $t_2$, which possesses a critical value of $v^2/24$. When $0<t_2<v^2/24$, the effective
potentials associated with the KK modes of the different $q$-form fields takes the form of
volcano-like potentials, which do not support localized massive modes. Nevertheless, massive
KK modes could appear as resonant states on the brane in this case. When $t_2=v^2/24$, the
effective potentials approach a positive constant when far away from the brane. Consequently,
a finite number of massive KK modes can be localized on the brane, and the number of localized
modes increases with the coupling parameter $t_1$. Finally, for $t_2>v^2/24$, the effective
potentials become infinitely deep potential wells. As a result, all massive modes of these
$q$-form fields can be localized on the brane, leading to the infinitely discrete spectra of
mass.

The paper is structured as follows: In section \ref{method}, we briefly review the 6D brane
background and introduce the curvature-dependent coupling between the $q$-form fields and the
gravity. The localization of various $q$-form fields is analyzed in section \ref{Loc}.
Specifically, the $0$-form scalar field, the $1$-form $U(1)$ gauge vector field, and the
$2$-form KR field are studied in sections \ref{scalar}, \ref{vector}, and \ref{KR}.
Finally, the main results and conclusions are summarized in section \ref{Cons}.

%%%%%%%%%%%%%%%%%%%%%%%%%%%%%%%%%%%%%%%%%%%%%%%%%%%%%

\section{The {  Method} }\label{method}

%To investigate the localization properties of the $q$-form fields in this 6D brane model, we introduce
%a nonminimal coupling between the $q$-form fields and the gravity. Such a coupling can be regarded as
%a generalization of the minimal coupling and may encode the effects of the background geometry on the
%dynamics of bulk fields. Since the spacetime curvature varies along the extra dimensions, the
%curvature-dependent coupling can significantly affect the localization behavior of the bulk fields.
%Based on this scenario, the action of a free 6D $q$-form field is taken to be

For a 6D $q$-form field $X_{M_1...M_q}$, an additional coupling will facilitate its localization on
the brane, which is usually introduced by adding a term or a multiplicative factor. The former
coupling method will lead to the breaking of the gauge invariance and necessitate further adjustments.
In contrary, the latter coupling method is free from this problem, and also have an effect on the
localization properties. Concerning this $q$-form field, we consider the latter coupling method and
suggest its action as
\begin{eqnarray} \label{actionq}
  S_q=\int d^6x\sqrt{-g}F(\phi,R,R^{\mu\nu}R_{\mu\nu},\cdots)Y_{M_1M_2...M_{q+1}}Y^{M_1M_2...M_{q+1}},
\end{eqnarray}
where $Y_{M_1M_2...M_{q+1}}=\partial_{[M_1}X_{M_2...M_{q+1}]}$ is the field strength tensor of the
$q$-form field $X_{M_1...M_q}$, and $F(\phi,R,R^{\mu\nu}R_{\mu\nu},\cdots)$ represents a scalar
function, which depends on the background scalar field $\phi$, the scalar curvature $R$, and other
geometric scalars. Such coupling can be regarded as a generalization of the minimal coupling and
may encode the effects of the various components of the bulk on the dynamics of bulk fields. This 
coupling function could be in the form of the geometry coupling. Except for what was mentioned above, the   
geometry coupling is also introduced in the localization of the $U(1)$ gauge fields \cite{ZHZhao2212.00444}, 
where the authors modify the original action by dropping the term $F_{AB}F^{AB}$ and adding a new term 
$\gamma_1R^{AB}_{\ \ \ CD}F_{AB}F^{CD}$, and obtain a localized zero-mass mode. In
ref. \cite{LYT2510.16491}, it is shown that the KR field can be localized on the brane when it is
coupled to the background scalar field. So, here we will focus on the coupling between the
$q$-form fields and the scalar curvature $R$, under which we express $F(\phi,R,R^{\mu\nu}R_{\mu\nu},\cdots)=F(R)$.

In this work, we consider the 6D spacetime as $\mathcal{M}_4\times\mathcal{R}_1\times\mathcal{S}_1$,
where $\mathcal{M}_4$ is the $3$-brane and $\mathcal{R}_1\times\mathcal{S}_1$ is the transverse
manifold. The line element of this 6D spacetime is
\begin{eqnarray} \label{6Dmetric0}
ds^2=a^2(y)\eta_{\mu\nu}dx^{\mu}dx^{\nu}+dy^2+b^2(y)R_0^2d\theta^2,
\end{eqnarray}
where $y\in(-\infty,\infty)$ denotes the large extra dimension and $\theta\in[0,2\pi)$ denotes
the compact extra dimension with radius $R_0$. Similar to the KK theory, we assume that the
radius $R_0$ of the compact extra dimension is sufficiently small such that it cannot be probed
at current experimental energy scales (of order TeV). By performing the coordinate transformation
$\Theta=R_0\theta$, the above line element can be rewritten as
\begin{eqnarray} \label{6Dmetricy}
ds^2=a^2(y)\eta_{\mu\nu}dx^{\mu}dx^{\nu}+dy^2+b^2(y)d\Theta^2.
\end{eqnarray}
Here the warp factors $a(y)$ and $b(y)$ are functions of the large extra dimension $y$, and
$\eta_{\mu\nu}$ is the metric of the $\mathcal{M}_4$ brane.

For this 6D spacetime, a brane model is proposed in ref. \cite{WJJjhep2105017}:
\begin{eqnarray}
\phi(y)&=&  v\ \text{sech}(ky),                                   \label{WarpFactor1}                    \\
   a(y)&=&  b(y)=e^{\frac{1}{24}v^2\tanh^2(ky)}\text{sech}^{\frac{v^2}{12}}(ky).     \label{WarpFactor2}
\end{eqnarray}
Here, $\phi$ is the background scalar field, $v$ is a dimensionless parameter, and $k$ is a
fundamental energy scale with dimension $[k]=L^{-1}$. It is straightforward to verify that
this brane model possesses $\mathbb{Z}_2$ symmetry. Therefore, we only consider the asymptotic
behavior of $a(y)$ in the limit $y\rightarrow+\infty$. The same convention will be adopted
for other quantities discussed below.

Furthermore, for this brane model, the scalar curvature of the bulk is
\begin{eqnarray} \label{ScaCurv}
  R&=&-10\frac{a''}{a}-20\frac{a'^2}{a^2}                            \nonumber          \\
   &=&\frac{5}{192}k^2v^2\big[48-3v^2+4(v^2+12)\cosh(2ky)-v^2\cosh(4ky)\big]                \nonumber          \\
   \vspace{0.2cm}
   & &\times\ \text{sech}^4(ky)\tanh^2(ky).
\end{eqnarray}
If $y\rightarrow+\infty$, its asymptotic solution can be obtained as
\begin{eqnarray} \label{AsymScaCurv}
  R\rightarrow-\frac{5}{24}k^2v^4+\frac{5}{6}k^2v^2(v^2+12)e^{-2ky}.
\end{eqnarray}
Thus, this brane model (\ref{WarpFactor2}) is asymptotically Anti-de Sitter. Besides, the stability of
this brane model has been demonstrated in ref. \cite{WJJjhep2105017}, and the corresponding $3$-brane
is located at the origin of the large extra dimension.

The function $F(R)$ should obey the following rules:
\begin{enumerate}
\item The coupling function $F(R)$ should be nonsingular.
\item If the scalar curvature $R \to 0$, the bulk spacetime becomes flat. The coupling should turn into
      the minimal coupling with $F(R)\rightarrow1$, in order that the action (\ref{actionq}) returns to
      the canonical one:
      \begin{equation} \label{action0}
        S_q=\int d^6x\sqrt{-g}Y_{M_1M_2...M_{q+1}}Y^{M_1M_2...M_{q+1}}.
      \end{equation}
\item { The function $F(R)$ should satisfy the positivity condition
      \begin{equation}
        F(R) >0 \label{positivity}
      \end{equation}
      to preserve the canonical form of 4D action}.
\end{enumerate}

For the specific form of the coupling function $F(R)$, both polynomial and exponential functions may be
considered. In ref. \cite{ZZHjhep1805072}, three different forms of $F(R)$ were proposed, two of which
are polynomial functions, while the third has an exponential form. Motivated by these studies, we focus on
an exponential curvature-dependent coupling in this work.

Furthermore, in dilaton-brane models, the coupling between matter fields and the dilaton field
$\pi$ is often introduced through an exponential factor of the form $e^{\xi\pi}$ \cite{Kehagias0010112,Fu1101.0336}.
Since the scalar curvature $R$ (\ref{ScaCurv}) given by is an even-parity quantity, analogous to
the dilaton field, it is natural to consider an exponential function of $R$ as the coupling factor
between the $q$-form fields and the gravity. Accordingly, in light of the scalar curvature (\ref{ScaCurv}),
we adopt the following form of the coupling function:
\begin{eqnarray} \label{FR}
  F(R)=e^{t_1\big(1-\big(\big(1+\frac{R}{\text C_R}\big)^2\big)^{-t_2/2}\big)},
\end{eqnarray}
where $t_1$ and $t_2$ are positive coupling parameters, and $\text C_R=\frac{5}{24}k^2v^4$. As will
be shown below, the parameter $t_2$ plays a crucial role in determining the structure of the effective
potentials appearing in the Schr\"{o}dinger-like equations for the KK modes.

%%%%%%%%%%%%%%%%%%%%%%%%%%%%%%
\section{Various $q$-form Fields}\label{Loc}

In this section, we study the localization of the KK modes for various $q$-form fields within
the 6D brane model (\ref{WarpFactor2}). In particular, we consider the $0$-form scalar field,
the $1$-form $U(1)$ gauge vector field, and the $2$-form KR field, and derive their corresponding
4D effective descriptions. Throughout this work, we have implicitly assumed that the energy
density carried by the bulk $q$-form fields is sufficiently small that their backreaction on
the background geometry can be neglected, so that the brane model (\ref{WarpFactor2}) remains
valid even in the presence of these bulk fields.

For the 6D brane model, we introduce the following coordinate transformation
\begin{eqnarray} \label{CoordTrans}
  \left\{
    \begin{array}{ll}
      dz=a^{-1}dy \\
      z=\int a^{-1}dy
    \end{array}
  \right.
\end{eqnarray}
with the boundary condition $z(y=0)=0$, where $z$ is the conformal coordinate of $y$. Then, based on
the brane model (\ref{WarpFactor2}), the line element (\ref{6Dmetricy}) becomes
\begin{eqnarray} \label{6Dmetricz}
  ds^2=a^2(z)(\eta_{\mu\nu}dx^{\mu}dx^{\nu}+dz^2+d\Theta^2).
\end{eqnarray}
In terms of this line element, the corresponding Schr\"{o}dinger-like equation of the KK modes
can be obtained, and various KK modes can be solved analytically or numerically.

Moreover, in the process of solving the KK modes, one finds that massive KK modes may exist as
resonant states on the brane when the corresponding effective potential has a volcano-like profile.
These resonant KK modes can be studied using the relative probability method proposed in
refs. \cite{LYX0904.1785,LYX0907.0910}. This method defines the relative probability of a resonance as
\begin{eqnarray} \label{PReso}
  P(m^2)=\frac{\int^{z_{\text b}}_{-z_{\text b}}|\psi(z)|^2dz}{\int^{z_{\text{max}}}_{-z_{\text{max}}}|\psi(z)|^2dz},
\end{eqnarray}
where $2z_{\text b}$ denotes the width of the brane, and $z_{\text{max}}$ is chosen to be $10z_{\text b}$.
For KK modes $\psi(z)$ whose squared masses $m^2$ are sufficiently larger than the maximum of the
corresponding effective potential, the wave functions approach plane-wave solutions. Consequently,
the relative probability approaches the value $0.1$, indicating that these modes are delocalized
in the extra dimension. The lifetime $\tau$ of a resonant state is estimated by $\tau\sim\Gamma^{-1}$,
where $\Gamma=\delta m$ denotes the full width at half maximum of the resonance peak. In the following,
we will focus on the localization of the KK modes of specific $q$-form fields, and figure out their
possible resonant modes using the above method.

%%%%%%%%%%%%%%%%%%%%%%%%%%%%%%%%%%%%%%%%%%
\subsection{Scalar fields}\label{scalar}

The action for a 6D massless scalar field is
\begin{eqnarray} \label{actionSca}
  S_0=-\frac12\int d^6x\sqrt{-g}\ g^{MN}F(R)\partial_M\Phi^*\partial_N\Phi.
\end{eqnarray}
Based on the metric (\ref{6Dmetricz}), performing the KK decomposition
\begin{eqnarray} \label{decompositionSca}
  \Phi(x^{\sigma},z,\Theta)=\sum_{m,n}\phi^{(m,n)}(x^{\sigma})\varphi_{(m,n)}(z)e^{il_n\Theta}a^{-2}F(R)^{-\frac12},
\end{eqnarray}
the action (\ref{actionSca}) can be written as
\begin{eqnarray} \label{actionRec}
  S_0&=&\sum_{m,n}-\frac12\bigg[I_{1(m,n)}\int d^4x\ \partial^{\mu}\phi^{*(m,n)}(x^{\sigma})
           \partial_{\mu}\phi^{(m,n)}(x^{\sigma})                                             \nonumber    \\
     & &+I_{2(m,n)}\int d^4x\ \phi^{*(m,n)}(x^{\sigma})\phi^{(m,n)}(x^{\sigma})\bigg],
\end{eqnarray}
where
\begin{eqnarray} \label{CoefSca}
   I_{1(m,n)}&=& 2\pi\int dz\ \varphi^*_{(m,n)}(z)\varphi_{(m,n)}(z),                                         \\
   I_{2(m,n)}&=& 2\pi\int dz\ a^4F(R)\partial_z\left[\varphi^*_{(m,n)}(z)a^{-2}F(R)^{-\frac12}\right]
                    \partial_z\left[\varphi_{(m,n)}(z)a^{-2}F(R)^{-\frac12}\right]                \nonumber    \\
             & & +2\pi l_n^2\int dz\ \varphi^*_{(m,n)}(z)\varphi_{(m,n)}(z).
\end{eqnarray}
It can be seen that $I_{2(m,n)}$ denotes the mass of the scalar field $\phi_{(m,n)}(x^{\sigma})$ in the
4D effective theory. The localization of the scalar zero mode requires
\begin{eqnarray} \label{LocScaZM}
  I_{1(0,0)}<\infty.
\end{eqnarray}

Varying the 6D action (\ref{actionSca}) with respect to $\Phi$, we obtain the equation of motion
\begin{eqnarray} \label{MoESca}
  \partial_M\left(\sqrt{-g}F(R)\partial^M\Phi\right)=0.
\end{eqnarray}
In terms of the KK decomposition (\ref{decompositionSca}), $\varphi_{(m,n)}(z)$ satisfies the following
Schr\"{o}dinger-like equation:
\begin{eqnarray} \label{SchroSca}
  \left[-\partial^2_z+U_0(z)\right]\varphi_{(m,n)}(z)=(m^2-l_n^2)\varphi_{(m,n)}(z),
\end{eqnarray}
where the effective potential $U_0(z)$ is
\begin{eqnarray} \label{EffPotSca}
  U_0(z)=\frac{2a''}{a}+\frac{2a'^2}{a^2}+\frac{F''(R)}{2F(R)}+\frac{2a'F'(R)}{aF(R)}-\frac{F'(R)^2}{4F(R)^2}.
\end{eqnarray}
This equation is defined along the large extra dimension, and the scalar KK modes $\varphi_{(m,n)}(z)$
likewise depend only on this coordinate. Since the 6D scalar field is automatically localized with
respect to the compact extra dimension, the localization analysis can be reduced to the study of
the KK modes $\varphi_{(m,n)}(z)$ associated with the large extra dimension.

Furthermore, the Schr\"{o}dinger-like equation (\ref{SchroSca}) can be factorized into
\begin{eqnarray} \label{FacSchroSca}
   \big(\partial_z+\Gamma'(z)\big)\big(-\partial_z+\Gamma'(z)\big)\varphi_{(m,n)}(z)
         =m_0^2\varphi_{(m,n)}(z),
\end{eqnarray}
where
\begin{eqnarray} \label{GammaSca}
  \Gamma'(z)=\frac{2a'}{a}+\frac{F'(R)}{2F(R)}.
\end{eqnarray}
For simplicity, we define $m_0^2=m^2-{l_n^2}$. The above equation can then be rewritten in the
supersymmetric form $B^{\dagger}B\varphi_{(m,n)}=m_0^2\varphi_{(m,n)}$ with $B=-\partial_z+\Gamma'$.
Since the operator $B^{\dagger}B$ is positive semi-definite, its eigenvalues satisfy $m_0^2\geq0$.
Therefore, no tachyonic modes are present, and the KK spectrum is stable.

\begin{figure} %[htb]
\begin{center}
\subfigure[$U_0(z).$] {\label{FigEffPotSca}
\includegraphics[width= 0.45\textwidth]{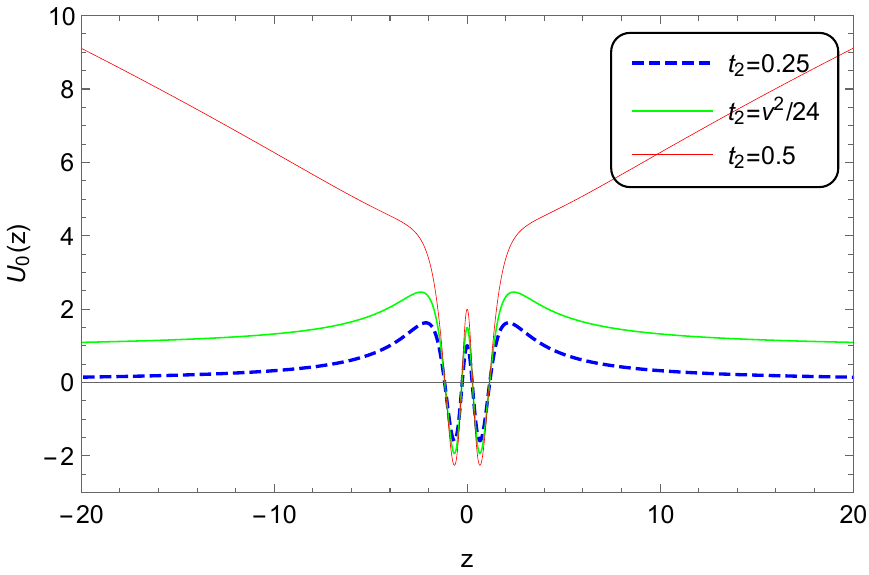}}
\subfigure[$\varphi_{(0,0)}(z).$] {\label{FigZMSca}
\includegraphics[width= 0.45\textwidth]{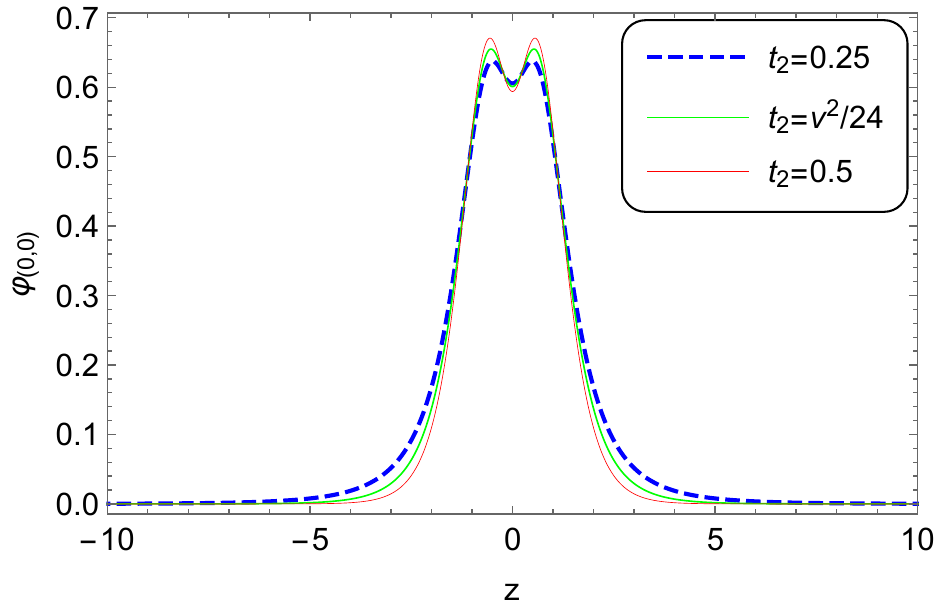}}
\end{center}\vskip -2mm
\caption{The effective potential $U_0(z)$ and the zero mode
         $\varphi_{(0,0)}(z)$ with the parameters $k=1$, $v=3$, $t_1=3$ and $t_2=0.25,v^2/24,0.5$.}
 \label{FigEffPotZMSca}
\end{figure}

This equation (\ref{FacSchroSca}) can give rise to the zero mode solution
\begin{eqnarray} \label{ZMSca}
  \varphi_{(0,0)}(z)=N_0a^2F(R)^{\frac12},
\end{eqnarray}
where $N_0$ is a constant. In terms of the coordinate transformation (\ref{CoordTrans}), the normalization
condition of this zero mode is
\begin{eqnarray} \label{NormContSca}
  \int|\varphi_{(0,0)}(z)|^2dz&=& \int|\varphi_{(0,0)}(z(y))|^2a^{-1}dy     \nonumber      \\
                              &=& N_0^2\int a^3F(R)dy=1.
\end{eqnarray}
Substituting eqs. (\ref{WarpFactor2}), (\ref{ScaCurv}) and (\ref{FR}) into this condition, we can obtain
\begin{eqnarray} \label{ExNormContSca}
  N_0^2\int\text{sech}^{\frac{v^2}{4}}(ky)
       e^{t_1-t_18^{t_2}\left(\frac{(5v^2-12+3(v^2+4)\cosh(4ky))^2\text{sech}^{12}(ky)}{v^4}\right)^{-t_2/2}
       +\frac18v^2\tanh^2(ky)} dy=1.
\end{eqnarray}
From this expression, it can be seen that the normalization condition (\ref{NormContSca}) is
satisfied provided that both parameters $t_1$ and $t_2$ are positive. Consequently, the zero mode $\varphi_{(0,0)}$
is normalizable, and the scalar field can be localized on the brane. The profiles of the this
zero mode $\varphi_{(0,0)}(z)$ are shown in fig. \ref{FigZMSca} with specific values of the
parameters.

For the massive KK modes, the localization behavior is governed by the effective potential (\ref{EffPotSca}).
In the context of the brane model (\ref{WarpFactor2}), however, this potential cannot be expressed
analytically in terms of the conformal coordinate $z$. To facilitate the analysis, we make use of
the coordinate transformation (\ref{CoordTrans}) and rewrite the effective potential in terms of
the coordinate $y$
\begin{eqnarray} \label{zyEffPotSca}
  U_0(z(y))=2aa''+4a'^2+\frac{a^2F''(R)}{2F(R)}+\frac{5aa'F'(R)}{2F(R)}-\frac{a^2F'(R)^2}{4F(R)^2},
\end{eqnarray}
where the prime denotes the derivative with respect to coordinate $y$. Substituting eqs. (\ref{WarpFactor2}),
(\ref{ScaCurv}) and (\ref{FR}) into this expression, we can obtain its asymptotic behaviors
\begin{eqnarray} \label{AsmpEffPotSca}
U_0(z(y\rightarrow\pm\infty))&\rightarrow&
\left\{
  \begin{array}{ll}
    +\infty, \hspace{0.5cm}   &t_2>v^2/24,  \\
    \text C_0, \hspace{0.5cm}  & t_2=v^2/24,  \\
    0,        &0<t_2<v^2/24
  \end{array}
\right.
\end{eqnarray}
with the positive limit
\begin{eqnarray} \label{LmtPTSca}
  \text C_0=\frac{1}{64}3^{-2-\frac{v^2}{12}}e^{\frac{v^2}{12}}k^2t_1^2v^4\left(\frac{v^2+4}{v^2}\right)^{-\frac{v^2}{12}}.
\end{eqnarray}
From the above expression, it is evident that the effective potential exhibits distinct asymptotic
behaviors as the coupling parameter $t_2$ varies. Numerical profiles of the effective potential
$U_0(z)$ for selected parameter values are presented in fig. \ref{FigEffPotSca}. Depending on the
value of $t_2$, the potential can exhibit volcano-like shape, a PT profile, or an infinitely deep
well. These three types of potentials lead to different localization properties of the massive KK
modes, which will be discussed separately below.

We first consider the case $0<t_2<v^2/24$. In this parameter region, the effective potential has
a volcano-like profile and does not support localized massive bound states on the brane. However,
although the effective potential tends to zero asymptotically when far away from the brane, the
potential well around the brane can temporarily trap the massive modes, giving rise to resonant
modes.

\begin{figure}[htb]
\begin{center}
\subfigure[$t_1=25,t_2=0.125$.]{\label{FigResSpecVolSca}
\includegraphics[width= 0.38\textwidth]{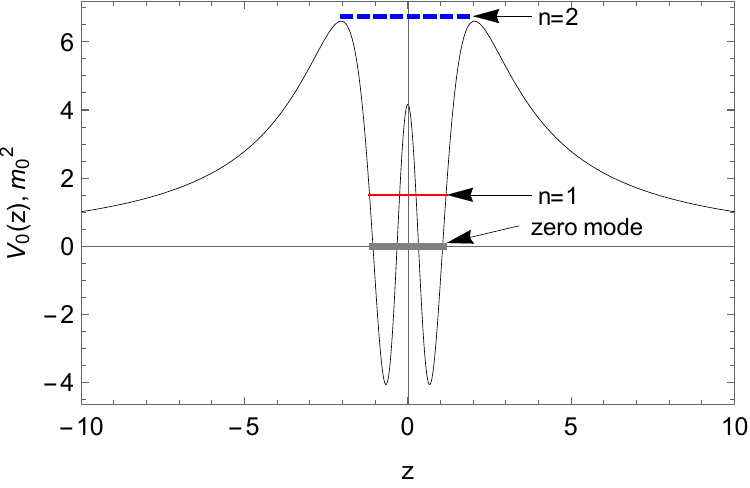}}
\hspace{0.5cm}
\subfigure[$t_1=25,t_2=0.125$.]{\label{FigResPVolSca}
\includegraphics[width= 0.38\textwidth]{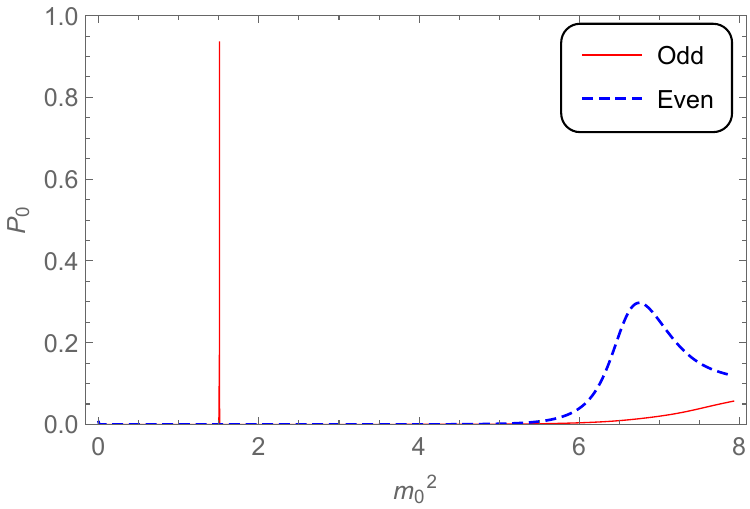}}
\subfigure[$t_1=20,t_2=0.25$.]{\label{2FigResSpecVolSca}
\includegraphics[width= 0.38\textwidth]{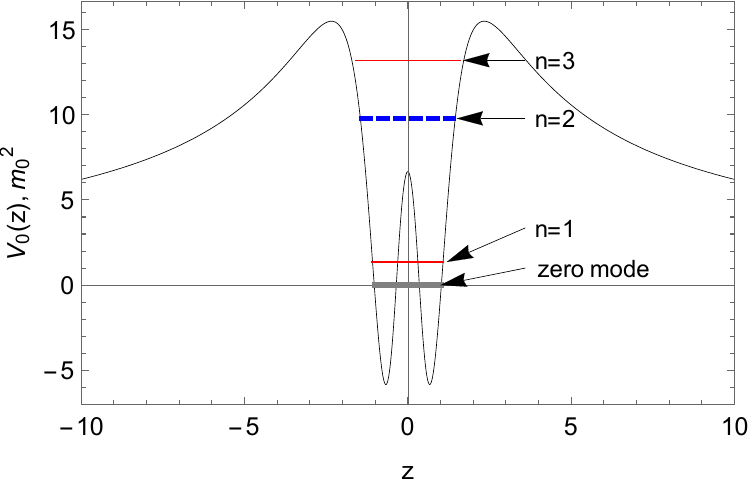}}
\hspace{0.5cm}
\subfigure[$t_1=20,t_2=0.25$.]{\label{2FigResPVolSca}
\includegraphics[width= 0.38\textwidth]{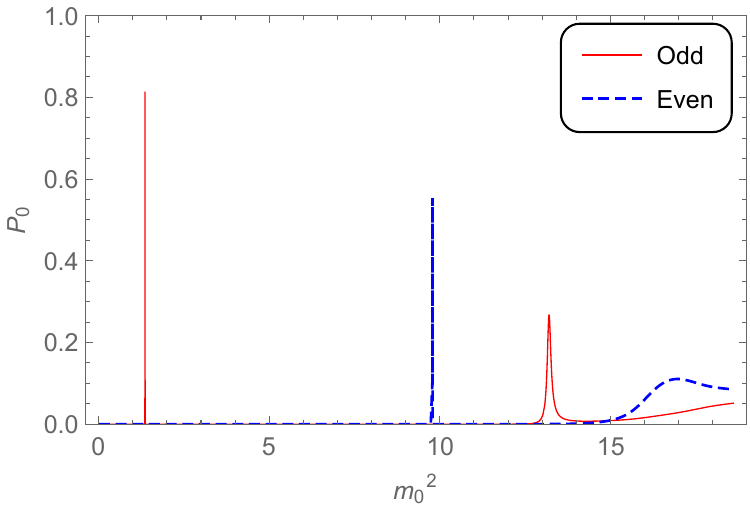}}
\subfigure[$t_1=25,t_2=0.25$.]{\label{3FigResSpecVolSca}
\includegraphics[width= 0.38\textwidth]{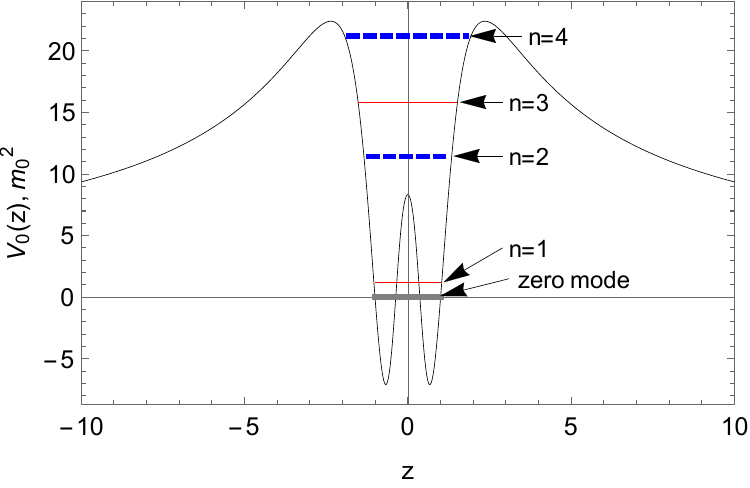}}
\hspace{0.5cm}
\subfigure[$t_1=25,t_2=0.25$.]{\label{3FigResPVolSca}
\includegraphics[width= 0.38\textwidth]{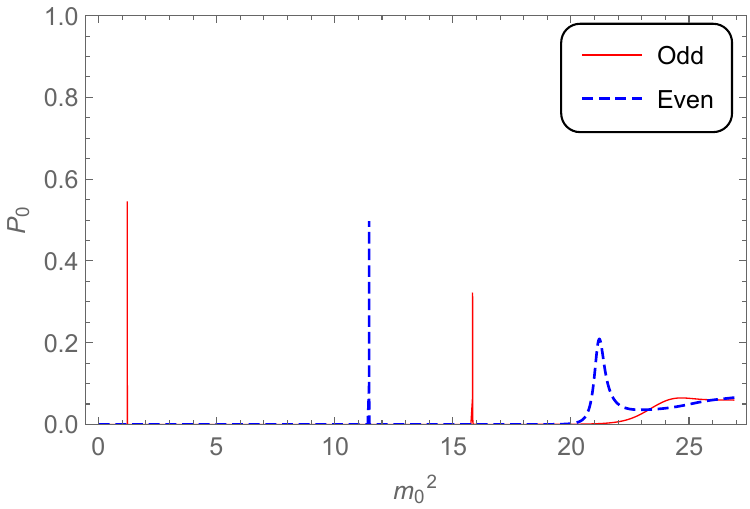}}
\end{center}\vskip -5mm
\caption{The mass spectra, the effective potential $U_0(z)$, and the corresponding relative probability
         $P_0$ with the parameters $k=1$, $v=3$ and $t=10,15,20$. The effective potential $U_0(z)$
         for the black line, the zero mode for the grey line, the even-parity resonant KK modes for
         the blue lines, and the odd-parity resonant KK modes for the red lines.}
 \label{SpecPResoVolSca}
\end{figure}

\begin{figure} %[htb]
\begin{center}
\subfigure[$\varphi$.]{\label{Odd1ResVolSca}
\includegraphics[width = 0.23\textwidth]{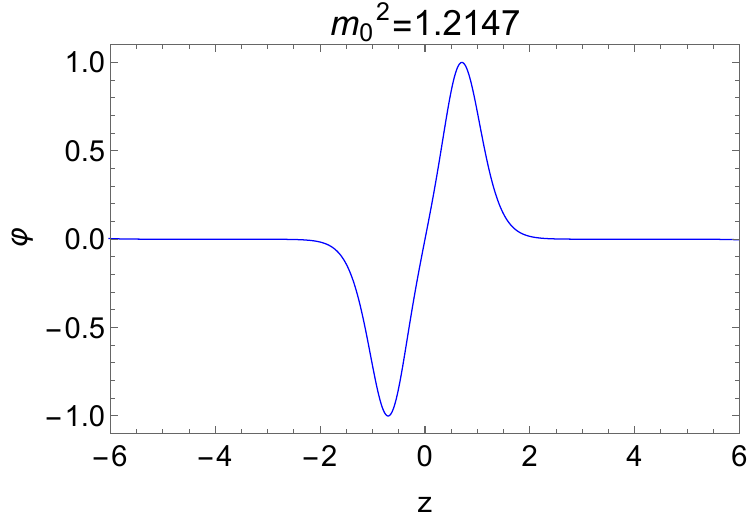}}
\subfigure[$\varphi$.]{\label{Even1ResVolSca}
\includegraphics[width = 0.23\textwidth]{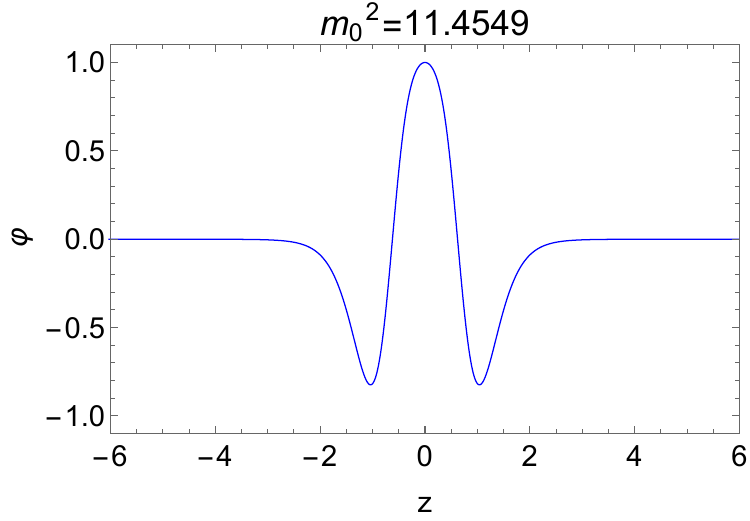}}
\subfigure[$\varphi$.]{\label{Odd2ResVolSca}
\includegraphics[width = 0.23\textwidth]{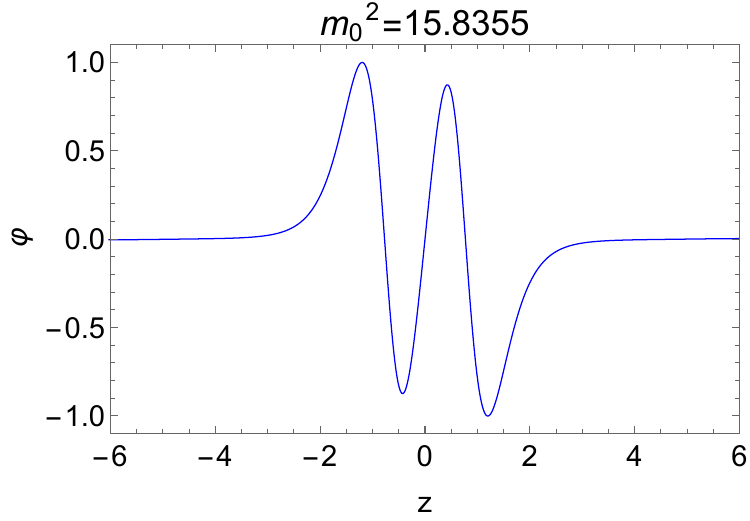}}
\subfigure[$\varphi$.]{\label{Even2ResVolSca}
\includegraphics[width = 0.23\textwidth]{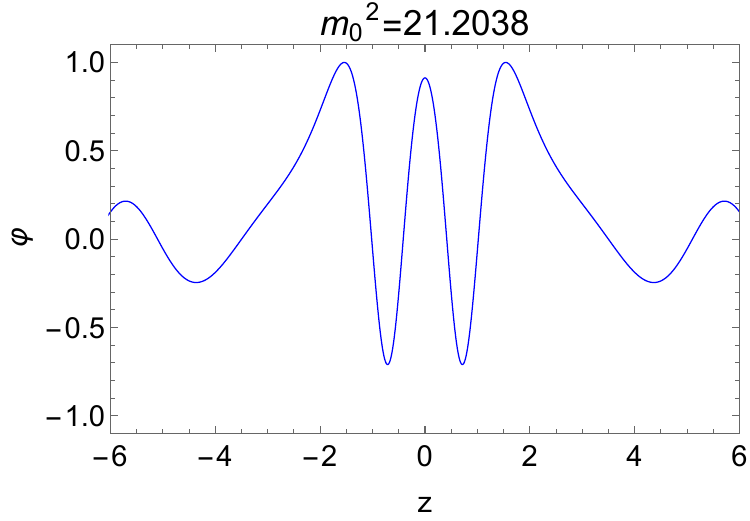}}
\end{center}\vskip -2mm
\caption{The shapes of resonant KK modes $\varphi(z)$ with different $m_0^{2}$. The parameters
         are set as $k=1,v=3,t_1=25$ and $t_{2}=0.25$.}
 \label{FigResVolSca}
\end{figure}

Using the relative probability method introduced above, we numerically solve for the resonant KK
modes associated with the effective potential $U_0(z)$ for various parameter values. The corresponding
relative probabilities and mass spectra are displayed in fig. \ref{SpecPResoVolSca}. In the mass
spectra, the zero mode corresponds to the ground state (bound state), and all massive KK modes appear
as resonant states. The masses, widths, and the lifetimes of the resonant KK modes are listed in
table \ref{tableRMOriSca}. It is found that the number of resonant KK modes increases with both
coupling parameters $t_1$ and $t_2$. As an illustrative example, the wave functions of four resonant
modes for $t_1=25$ and $t_2=0.25$ are shown in fig. \ref{FigResVolSca}. Therefore, for $0<t_2<v^2/24$,
a finite number of the massive KK modes can be quasi-localized on the brane and manifest themselves
as resonant states.
%%%%%%%%%%%%%%%%%%%%%%%%%%%%%%%%%%%%%%%%%%%%%%%%%%%%%%%%%%%%%%%%%%%%%%%%%%%%
\begin{table}[tbp]
\centering
\begin{tabular}{|c|c|c|c|c|c|c|c|}
    \hline
    $t_1$               & $t_2$                 & $U^{\text{max}}_{0}$  & $n$         & $m_0^2$ &
    $m_0$               & $\Gamma$              & $\tau$
    \\
    \hline
    $25$                & $0.125$               & $6.6054$              & $1$         & $1.5083$   &
    $1.2281$            & $6.839\times10^{-5}$  & $1.462\times10^{4}$
    \\
                        &                       &                       & $2$         & $6.7522$ &
    $2.5985$            & $0.2176$              & $4.5949$
    \\
    \hline
    $20$                & $0.25$                & $15.5016$             & $1$         & $1.3605$ &
    $1.1664$            & $3.771\times10^{-11}$ & $2.652\times10^{10}$
    \\
    %\cline{3-9}
                        &                       &                       & $2$         & $9.7858$ &
    $3.1282$            & $2.177\times10^{-4}$  & $4.593\times10^3$
    \\
                        &                       &                       & $3$         & $13.2021$ &
    $3.6335$            & $0.0184$              & $54.2805$
    \\
    \hline
    $25$                & $0.25$                & $22.4093$             & $1$         & $1.2147$ &
    $1.1021$            &$1.611\times10^{-10}$  & $6.206\times10^9$
    \\
                        &                       &                       & $2$         & $11.4549$ &
    $3.3845$            & $1.112\times10^{-6}$  & $8.993\times10^5$
    \\
                        &                       &                       & $3$         & $15.8355$ &
    $3.9794$            & $4.619\times10^{-4}$  & $2.165\times10^3$
    \\
                        &                       &                       & $4$         & $21.2038$ &
    $4.6048$            & $0.622$               & $16.079$
    \\
    \hline
\end{tabular}
\caption{The masses, widths, and lifetimes of resonant KK modes $\varphi(z)$.
         The parameters are set as $k=1$ and $v=3$. }
    \label{tableRMOriSca}
\end{table}
%%%%%%%%%%%%%%%%%%%%%%%%%%%%%%%%%%%%%%%%%%%%%%%%%%%%%%%%%%%%%%%%%%%%%%%%%%%%%%%%%%%%%

\begin{figure} %[htbp]
\begin{center}
\includegraphics[width= 0.49\textwidth]{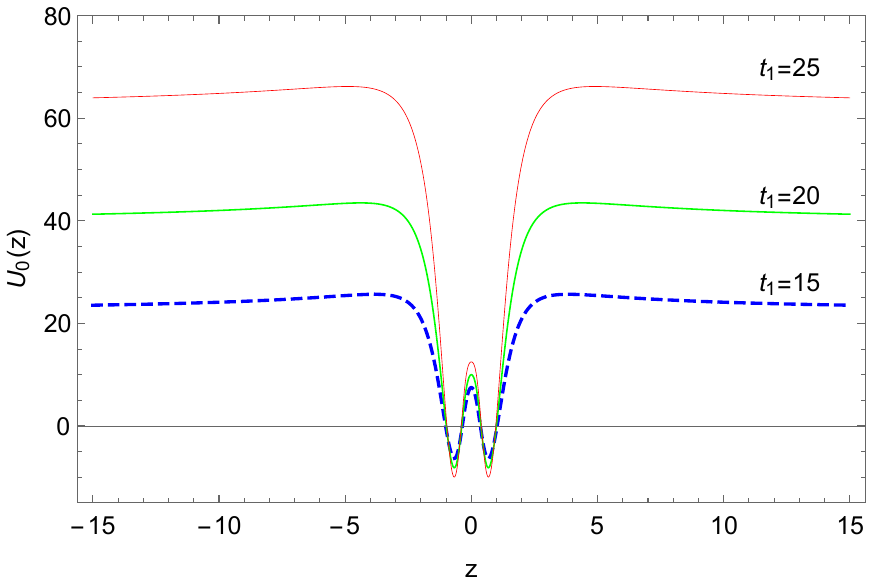}
\caption{The effective potential $U_0(z)$ with the parameter $t_1=15,20$, and $25$.
        The other parameters are set as $k=1,v=3$, and $t_2=3/8$.}
\label{figSpecPTSca}
\end{center}
\end{figure}

Next, we consider the case $t_2=3/8$, which corresponds to the critical value $t_2=v^2/24$. In this
case, the effective potential takes the form of a PT potential, which can support a finite number
of bound states. According to eq. (\ref{LmtPTSca}), the depth of the potential well depends on the
parameters $v$, $k$, and $t_1$. Consequently, the number of localized massive KK modes is also
determined by these parameters. Since our primary interest is the effect of the coupling mechanism,
characterized by the parameters $t_1$ and $t_2$, we focus on the influence of these two parameters.
To this end, the effective potentials for $t_1=15$, $20$, and $25$ are plotted in fig. \ref{figSpecPTSca}
with other parameters $k=1$, $v=3$, and $t_2=3/8$. As shown in the figure, the potential well
becomes deeper as $t_1$ increases. For these three effective potentials, the localized massive
KK modes are obtained numerically, and the corresponding mass spectra are
\begin{eqnarray}
  m_0^2=&\hspace{-3.71cm}\{0,1.30,10.61,14.79,20.24\},                     &\text{for}~t_1=15;    \label{SpecPTSca1} \\
  m_0^2=&\hspace{-1.63cm}\{0,1.08,12.96,18.12,25.78,31.84,37.00\},         &\text{for}~t_1=20;    \label{SpecPTSca2} \\
  m_0^2=&\{0,0.86,15.27,30.45,38.75,46.32,52.97,58.26\},\hspace{0.6cm}     &\text{for}~t_1=25.     \label{SpecPTSca3}
\end{eqnarray}
Therefore, when $t_2=v^2/24$, the effective potential is of the PT type and supports a finite number
of localized massive KK modes. Furthermore, the number of localized massive modes increases with the
coupling parameter $t_1$.

\begin{figure} %[htbp]
\begin{center}
\includegraphics[width= 0.49\textwidth]{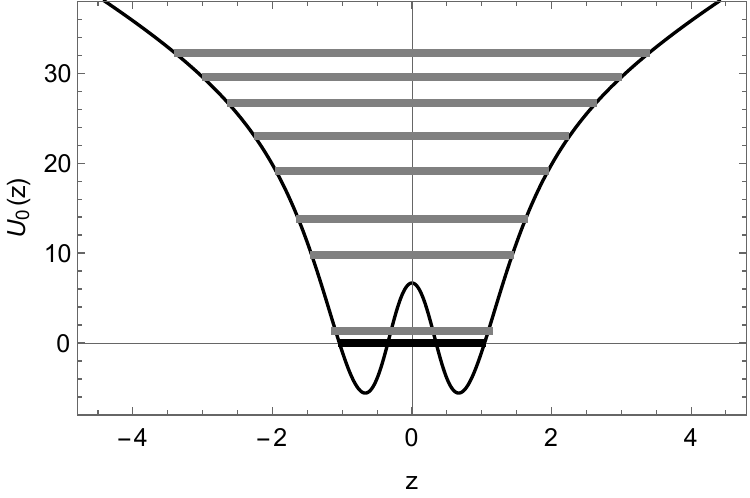}
\caption{The effective potential $U_0(z)$ with the parameters $k=1,v=3,t_1=10$, and $t_2=0.5$.}
\label{figSpecIDWSca}
\end{center}
\end{figure}

Finally, we consider the case $t_2=0.5$, which is greater than the critical value $v^2/24$. In this
regime, the effective potential becomes an infinitely deep potential well. Consequently, all massive
KK modes are localized on the brane, giving rise to an infinitely discrete mass spectrum. The corresponding
mass spectrum, including the zero mode and the low-lying massive KK modes, is shown in fig. \ref{figSpecIDWSca}
for the parameter values $k=1,v=3$ and $t_1=10$.

Therefore, the scalar field can be localized on the thick brane in the 6D spacetime through its
coupling to the gravity. The localization properties of the massive KK modes are determined by
the coupling parameter $t_2$ and can be classified into three distinct cases. For $0<t_2<v^2/24$,
the effective potential has a volcano-like shape and does not support localized massive bound
modes. Nevertheless, massive KK modes may appear as resonant states on the brane, and the number
of such resonances increases with both $t_1$ and $t_2$. At the critical value $t_2=v^2/24$, the
effective potential takes the form of a PT potential, whose asymptotic value is proportional to
$t_1^2$. In this case, a finite number of massive KK modes can be localized on the brane, and
the number of localized modes increases with the coupling parameter $t_1$. For $t_2>v^2/24$, the
effective potential becomes an infinitely deep potential well. Consequently, all massive KK modes
are localized on the thick brane, resulting in an infinitely discrete spectrum of mass.

%%%%%%%%%%%%%%%%%%%%%%%%%%%%%%

\subsection{$U(1)$ gauge vector fields}\label{vector}

In this section, we investigate the localization of the 6D $U(1)$ gauge vector field when it is
coupled with the gravity. We follow the method in refs. \cite{WJJjhep2105017,FCEjhep1901021}
and choose a similar KK decomposition for this $U(1)$ gauge vector field. In ref. \cite{FCEjhep1901021},
the authors verified that the corresponding 4D effective action of the 6D $U(1)$ gauge vector
field is gauge invariant under the minimal coupling. Here, the coupling function $F(R)$ depends
only on the large extra dimension, and it can also be demonstrated that the 4D effective action
is gauge invariant.

Considering the coupling between the 6D $U(1)$ gauge vector field and the gravity, we start with the
action
\begin{eqnarray} \label{actionVec}
  S_1=-\frac14\int d^6x\sqrt{-g}F(R)F^{MN}F_{MN},
\end{eqnarray}
where the field strength tensor
\begin{eqnarray} \label{fieldStrenVec}
  F_{MN}=\partial_MA_N-\partial_NA_M.
\end{eqnarray}

Based on the metric (\ref{6Dmetricz}), we perform the following KK decomposition
\begin{eqnarray} \label{1decompositionVec}
  A_{\mu}    &=& \sum_m\hat X^{(m)}_{\mu}(x^{\sigma})W_1^{(m)}(z,\Theta)a^{-1},             \\
  A_z        &=& \sum_m\hat{\zeta}^{(m)}(x^{\sigma})W_2^{(m)}(z,\Theta)a^{-1},              \\
  A_{\Theta} &=& \sum_m\hat{\xi}^{(m)}(x^{\sigma})W_3^{(m)}(z,\Theta)a^{-1},
\end{eqnarray}
where $\hat X^{(m)}_{\mu}$ is the 4D vector field, $\hat{\zeta}^{(m)}(x^{\sigma})$ and $\hat{\xi}^{(m)}(x^{\sigma})$
are two 4D scalar fields. Then, the 6D action (\ref{actionVec}) can be reduced to the 4D effective one
\begin{eqnarray}\label{1actionVec}
  S_1&=&-\frac{1}{4}\int {d^6 x} \sqrt{-g}F(R)F^{MN}F_{MN}\nonumber\\
  &=&-\frac{1}{4}\int {d^6 x} \sqrt{-g}F(R)\bigg(F^{\mu_1\mu_2}F_{\mu_1\mu_2}+2F^{\mu_1z}F_{\mu_1z}
  +2F^{\mu_1\Theta}F_{\mu_1\Theta}+2F^{\Theta z}F_{\Theta z}\bigg)
  \nonumber\\
   &=&
     -\frac{1}{4}\sum_m\sum_{m'}\int {d^{4}x} \sqrt{-\hat{g}}
     \;\bigg[
            I_{1}^{(mm')}\;\hat{F}^{(m)}_{\mu_1\mu_2}\;\hat{F}^{\mu_1\mu_2(m')}
            +\big(I_{2}^{(mm')}+I_{4}^{(mm')}\big)\;\hat{X}_{\mu_1}^{(m)}\hat{X}^{\mu_1(m')}\nonumber\\
          &&+I_{3}^{(mm')}\;\partial_{\mu_1}\hat{\zeta}^{(m)}\;\partial^{\mu_1}\hat{\zeta}^{(m')}
            -I_{6}^{(mm')}\;\bigg(\partial_{\mu_1}\hat{\zeta}^{(m)}\;\hat{X}^{\mu_1(m')}
                                +\hat{X}_{\mu_1}^{(m)}\;\partial^{\mu_1}\hat{\zeta}^{(m')}\bigg)\nonumber\\
          &&+I_{5}^{(mm')}\;\partial_{\mu_1}\hat{\xi}^{(m)}\;\partial^{\mu_1}\hat{\xi}^{(m')}
            -I_{8}^{(mm')}\;\bigg(\partial_{\mu_1}\hat{\xi}^{(m)}\;\hat{X}^{\mu_1(m')}
                                +\hat{X}_{\mu_1}^{(m)}\;\partial^{\mu_1}\hat{\xi}^{(m')}\bigg)\nonumber\\
          &&+I_{7}^{(mm')}\;\hat{\zeta}^{(m)}\hat{\zeta}^{(m')}
            +I_{9}^{(mm')}\;\hat{\xi}^{(m)}\hat{\xi}^{(m')}
            -\;I_{10}^{(mm')}\;
            \bigg(\hat{\zeta}^{(m)}\hat{\xi}^{(m')}+\hat{\xi}^{(m)}\hat{\zeta}^{(m')}\bigg)
                                 \bigg],
\end{eqnarray}
where $\hat{F}^{(m)}_{\mu\nu}=\partial_{\mu}\hat X_{\nu}^{(m)}-\partial_{\nu}\hat X_{\mu}^{(m)}$ and the
constants are given by
\begin{subequations}\label{constantI}
\begin{eqnarray}
&&{I}_{1}^{(mm')}\equiv\int d \Theta d z~ F(R)W_{1}^{(m)} W_{1}^{(m')} ,           \\
&&{I}_{2}^{(mm')}\equiv2 \int {d} \Theta {d} z~ F(R)\partial_{\Theta}\left(W_{1}^{(m)} a^{-1}\right)
        \partial_{\Theta}\left({W}_{1}^{(m')} a^{-1}\right) a^{2}  ,           \\
&&{I}_{3}^{(mm')}\equiv2 \int {d} \Theta {d} z~ F(R)W_{2}^{(m)} {W}_{2}^{(m')} ,   \\
&&{I}_{4}^{(mm')}\equiv2 \int {d} \Theta {d} z~ F(R)\partial_{z}\left(W_{1}^{(m)} a^{-1}\right)
        \partial_{z}\left(W_{1}^{(m')} a^{-1}\right) a^{2} ,                   \\
&&{I}_{5}^{(mm')}\equiv2 \int {d\Theta} {d} z~ F(R){W}_{3}^{(m)} {W}_{3}^{(m')},   \\
&&{I}_{6}^{(mm')}\equiv2 \int d \Theta d z~ F(R)W_{2}^{(m)} \partial_{z}\left(W_{1}^{(m')} a^{-1}\right) a ,     \\
&&{I}_{7}^{(mm')}\equiv2 \int {d} \Theta {d} z~ F(R)\partial_{\Theta}\left({W}_{2}^{(m)} a^{-1}\right)
        \partial_{\Theta}\left({W}_{2}^{(m')} a^{-1}\right) a^{2},             \\
&&{I}_{8}^{(mm')}\equiv2 \int {d\Theta dz}~ F(R){W}_{3}^{(m)} \partial_{\Theta}\left({W}_{1}^{(m')} a^{-1}\right) a  ,\\
&&{I}_{9}^{(mm')}\equiv2 \int d \Theta d z~ F(R)\partial_{z}\left(W_{3}^{(m)} a^{-1}\right)
        \partial_{z}\left(W_{3}^{(m')} a^{-1}\right) a^{2},                    \\
&&{I}_{10}^{(mm')}\equiv2 \int {d} \Theta {d} z~ F(R)\partial_{z}\left(W_{3}^{(m)} a^{-1}\right)
        \partial_{\Theta}\left({W}_{2}^{(m')} a^{-1}\right) a^{2}.
\end{eqnarray}
\end{subequations}
By varying this 4D effective action (\ref{1actionVec}) with respect to $\hat X_{\mu}^{(m)}$,
$\hat{\zeta}^{(m)}$ and $\hat{\xi}^{(m)}$, we can get
\begin{subequations}\label{vary4DVec}
\begin{align}
&\frac{{I}_{1}^{(mm')} }{\sqrt{-\hat{g}}} \partial_{\mu_{1}}\left(\sqrt{-\hat{g}}~\hat{F}^{(m) \mu_{1} \mu_{2}}\right)
-({I}_{2}^{(mm')}+{I}_{4}^{(mm')})\hat{X}^{\mu_{2}(m)}
+{I}_{6}^{(mm')} \partial^{\mu_2} \hat{\zeta}
+{I}_{8}^{(mm')} \partial^{\mu_2}\hat{\xi}=0,\\
&\frac{I_{3}^{(mm')}}{\sqrt{-\hat{g}}}\partial_{\mu_1}
  \left(\sqrt{-\hat{g}}~\partial^{\mu_1}\hat{\zeta}^{(m')}\right)
  -\frac{I_{6}^{(mm')}}{\sqrt{-\hat{g}}}\partial_{\mu_1} \left(\sqrt{-\hat{g}}~\hat{X}^{\mu_1(m')}\right)
  -I_{7}^{(mm')}\hat{\zeta}^{(m')}+I_{10}^{(mm')}\hat{\xi}^{(m')}=0,\\
&\frac{I_{5}^{(mm')}}{\sqrt{-\hat{g}}}\partial_{\mu_1}
  \left(\sqrt{-\hat{g}}~\partial^{\mu_1}\hat{\xi}^{(m')}\right)
  -\frac{I_{8}^{(mm')}}{\sqrt{-\hat{g}}}\partial_{\mu_1} \left(\sqrt{-\hat{g}}~\hat{X}^{\mu_1(m')}\right)
  -I_{9}^{(mm')}\hat{\xi}^{(m')}+I_{10}^{(mm')}\hat{\zeta}^{(m')}=0.\label{effequ2}
\end{align}
\end{subequations}

On the other hand, by varying the 6D action (\ref{actionVec}) with respect to $A_M$, we can obtain the
equation of motion
\begin{eqnarray} \label{MoEVec}
  \frac{1}{\sqrt{-g}}\partial_M\left(\sqrt{-g}F(R)F^{MN}\right)=0.
\end{eqnarray}
This equation contains the component equations
\begin{subequations}\label{ComnEoMVec}
\begin{align}
&\frac{1}{\sqrt{-\hat{g}}}\partial_{\mu_1}\left(\sqrt{-\hat{g}}\hat{F}^{\mu_1\mu_2(m)}\right)
+(\lambda_{1} +\lambda_{2}) \hat{X}^{\mu_2(m)}
-\lambda_{4} \partial^{\mu_2} \hat{\zeta}^{(m)}
-\lambda_{3} \partial^{\mu_2} \hat{\xi}^{(m)}=0, \\
&\frac{1}{\sqrt{-\hat{g}}}\partial_{\mu_1}\left(\sqrt{-\hat{g}}~\hat{g}^{\mu_1\mu_2}\partial_{\mu_2}\hat{\zeta}^{(m)}\right)
-\lambda_{5}\frac{1}{\sqrt{-\hat{g}}}\partial_{\mu_1}\left(\sqrt{-\hat{g}}~\hat{g}^{\mu_1\mu_2}\hat{X}_{\mu_2}^{(m)}\right)
+\lambda_{6} \hat{\zeta}^{(m)}
-\lambda_{7} \hat{\xi}^{(m)}=0, \\
&\frac{1}{\sqrt{-\hat{g}}}\partial_{\mu_1}\left(\sqrt{-\hat{g}}~\hat{g}^{\mu_1\mu_2}\partial_{\mu_2}\hat{\xi}^{(m)}\right)
-\lambda_{8}\frac{1}{\sqrt{-\hat{g}}}\partial_{\mu_1}\left(\sqrt{-\hat{g}}~\hat{g}^{\mu_1\mu_2}\hat{X}_{\mu_2}^{(m)}\right)
-{\lambda_{9} \hat{\zeta}^{(m)}
+\lambda_{10} \hat{\xi}^{(m)}=0},
\end{align}
\end{subequations}
where
\begin{eqnarray}\label{lambdaVec}
\lambda_{1} &\equiv& \frac{\partial_{\Theta}\left(a^{2}F(R)\partial_{\Theta}\left(W_{1}^{(m)} a^{-1}\right)\right)
               }{aF(R)W_{1}^{(m)}},~~~~~~~
\lambda_{2}\equiv\frac{\partial_{z}\left(a^{2} F(R)\partial_{z}\left(W_{1}^{(m)} a^{-1}\right)\right)}{aF(R)W_{1}^{(m)}},\nonumber \\
\lambda_{3}&\equiv&\frac{\partial_{\Theta}\left(aF(R)W_{3}^{(m)}\right)}{aF(R)W_{1}^{(m)}},~~~~~~~~~~~~~~~~~~~~~
\lambda_{4}\equiv\frac{\partial_{z}\left(aF(R)W_{2}^{(m)}\right) }{aF(R)W_{1}^{(m)}},\nonumber\\
\lambda_{5}&\equiv&\frac{\partial_{z}\left(W_{1}^{(m)} a^{-1}\right) a}{W_{2}^{(m)}},~~~~~~~~~~~~~~~~~~~~~~~~
\lambda_{6}\equiv\frac{\partial_{\Theta}\left(\partial_{\Theta}\left(W_{2}^{(m)} a^{-1}\right) a^{2}\right) a^{-1}}{W_{2}^{(m)}}, \nonumber \\
\lambda_{7}&\equiv&\frac{\partial_{\Theta}\left(\partial_{z}\left(W_{3}^{(m)} a^{-1}\right) a^{2}\right) a^{-1}}{W_{2}^{(m)}},~~~~~~~~~\,
\lambda_{8}\equiv\frac{\partial_{\Theta}\left(W_{1}^{(m)} a^{-1}\right) a}{W_{3}^{(m)}},\nonumber\\
\lambda_{9}&\equiv&\frac{\partial_{z}\left(\partial_{\Theta}\left(W_{2}^{(m)} a^{-1}\right) a^{2}\right) a^{-1}}{W_{3}^{(m)}},~~~~~~~~\,
\lambda_{10}\equiv\frac{\partial_{z}\left(\partial_{z}\left(W_{3}^{(m)} a^{-1}\right) a^{2}\right) a^{-1}}{W_{3}^{(m)}}.
\end{eqnarray}

Both eqs. (\ref{vary4DVec}) and (\ref{lambdaVec}) are derived from the fundamental 6D action (\ref{actionVec}),
and they should be compatible, so there are conditions:
\begin{align}\label{consistency relationship}
&I_{1}^{(mm')}=\delta^{mm'},~~~~~~~~~I_{2}^{(mm')}=-\lambda_1\delta^{mm'},~~~
I_{4}^{(mm')}=-\lambda_2\delta^{mm'},~~~~I_{6}^{(mm')}=-\lambda_4\delta^{mm'},\nonumber\\
&I_{8}^{(mm')}=-\lambda_3\delta^{mm'},~~~~I_{3}^{(mm')}=\delta^{mm'},~~~~~~~
I_{6}^{(mm')}=\lambda_5\delta^{mm'},~~~~~~~~I_{7}^{(mm')}=-\lambda_6\delta^{mm'},\nonumber\\
&I_{10}^{(mm')}=-\lambda_7\delta^{mm'},~~~~I_{5}^{(mm')}=\delta^{mm'},~~~~~~~
I_{8}^{(mm')}=\lambda_8\delta^{mm'},~~~~~~~~I_{9}^{(mm')}=\lambda_{10}\delta^{mm'},\nonumber\\
&I_{10}^{(mm')}=\lambda_9\delta^{mm'}.
\end{align}
These conditions impose constraints on the higher-dimensional model to ensure that the
higher-dimensional theory is compatible with observations.

We further separate variables as
\begin{eqnarray}\label{2decompositionVec}
W_1^{(m)}(z,\Theta)&=&\sum_{n}w^{(m,n)}_1(z)F(R)^{-\frac12}e^{il_n\Theta},\\
W_2^{(m)}(z,\Theta)&=&\sum_{n}w^{(m,n)}_2(z)F(R)^{-\frac12}e^{il_n\Theta},\\
W_3^{(m)}(z,\Theta)&=&\sum_{n}w^{(m,n)}_3(z)F(R)^{-\frac12}e^{il_n\Theta},
\end{eqnarray}
where $w^{(m,n)}_1(z),w^{(m,n)}_2(z)$ and $w^{(m,n)}_3(z)$ are the KK modes of the 4D fields
$\hat X^{(m)}_{\mu}$, $\hat{\zeta}^{(m)}$ and $\hat{\xi}^{(m)}$. Then, eqs. (\ref{lambdaVec})
can be rewritten as
\begin{eqnarray}\label{lambdawVec}
\lambda_{1}&\equiv& -l_n^2,                                                             \hspace{4.73cm}
    \lambda_{2}w_1\equiv \frac{\partial_z\left(a^2F(R)\partial_z\left(w_1F(R)^{-\frac12}a^{-1}\right)\right)}
                     {aF(R)^{\frac12}},                                                           \nonumber \\
\lambda_{3}w_1&\equiv& il_nw_3,                                                         \hspace{4.5cm}
    \lambda_{4}w_1\equiv \frac{\partial_z\left(aF(R)^{\frac12}w_2\right)}{aF(R)^{\frac12}},       \nonumber\\
\lambda_{5}w_2&\equiv& \partial_z\left(w_1F(R)^{-\frac12}a^{-1}\right)aF(R)^{\frac12},                             \hspace{1.17cm}
    \lambda_{6}\equiv -l_n^2,                                                                     \nonumber\\
\lambda_{7}w_2&\equiv& il_n\partial_z\left(w_3F(R)^{-\frac12}a^{-1}\right)aF(R)^{\frac12},                         \hspace{0.25cm}
    \lambda_{8}w_3\equiv il_nw_1,                                                                 \nonumber\\
\lambda_{9}w_3&\equiv& \frac{il_n\partial_z\left(w_2F(R)^{-\frac12}a\right)}{aF(R)^{-\frac12}},                    \hspace{1.8cm}
    \lambda_{10}w_3\equiv \frac{\partial_z\left(\partial_z\left(w_3F(R)^{-\frac12}a^{-1}\right)a^2\right)}
                       {aF(R)^{-\frac12}}.
\end{eqnarray}

By canonically normalizing the action (\ref{1actionVec}), The 4D effective action can be written as
\begin{equation}\label{2actionVec}
S_1=-\frac{1}{4}\sum_m\sum_{m'}\int {d^{4}x} \sqrt{-\hat{g}}
     \;\bigg(
            \;\hat{F}^{(m)}_{\mu_1\mu_2}\;\hat{F}^{\mu_1\mu_2(m')}
            +\frac{I_{2}^{(mm')}+I_{4}^{(mm')}}{I_{1}^{(mm')}}\;\hat{X}_{\mu_1}^{(m)}\hat{X}^{\mu_1(m')}+\cdots\bigg).
\end{equation}
In light of the consistency conditions (\ref{consistency relationship}), there is
\begin{eqnarray}\label{Corres}
  \lambda_1+\lambda_2=-\frac{I_{2}^{(mm')}+I_{4}^{(mm')}}{I_{1}^{(mm')}},
\end{eqnarray}
and the mass $m$ is given by
\begin{eqnarray} \label{massVec}
  m^2=-(\lambda_1+\lambda_2).
\end{eqnarray}
Substituting $\lambda_1$ and $\lambda_2$ (\ref{lambdawVec}) into (\ref{massVec}), we can
obtain
\begin{equation}\label{SchroVec}
\left[ {-\partial_z^2 + U_1(z)} \right]w_1^{(m,n)} =\big(m^2-{l_n^2}\big)w_1^{(m,n)}~,
\end{equation}
where the effective potential $U_1(z)$ is
\begin{equation}\label{EffPotVec}
  U_1(z)=\frac{a''}{a}+\frac{F''(R)}{2F(R)}+\frac{a'F'(R)}{aF(R)}-\frac{F'(R)^2}{4F(R)^2}.
\end{equation}
This equation can further be factorised into
\begin{equation}\label{2SchroVec}
\left(\partial_z+\Gamma'(z)\right)\left( -\partial_z +\Gamma'(z)\right)w_1^{(m,n)} =m_1^2w_1^{(m,n)}~,
\end{equation}
with
\begin{eqnarray}\label{GammaVec}
  \Gamma'(z)=\frac{a'}{a}+\frac{F'(R)}{2F(R)},
\end{eqnarray}
where, for convenience, we have defined $m_1^2=m^2-{l_n^2}$. This equation is in the form of
$B^{\dagger}Bw_1^{(m,n)} =m_1^2w_1^{(m,n)}$ with $B= -\partial_z +\Gamma'(z)$, which
shows that $m_1^2\geq0$ and hence there is no tachyonic state.

\begin{figure} %[htb]
\begin{center}
\subfigure[$U_1(z).$] {\label{FigEffPotVec}
\includegraphics[width= 0.45\textwidth]{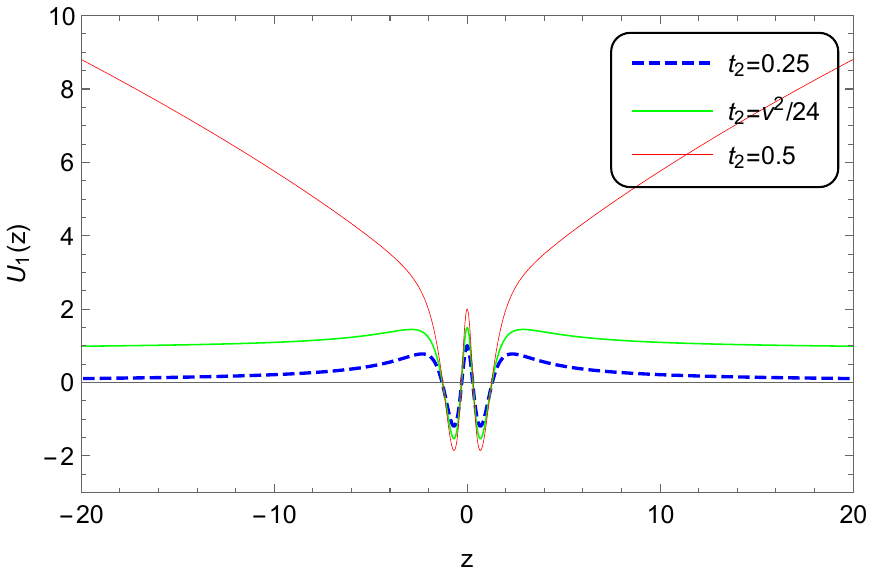}}
\subfigure[$w_1^{(0,0)}(z).$] {\label{FigZMVec}
\includegraphics[width= 0.45\textwidth]{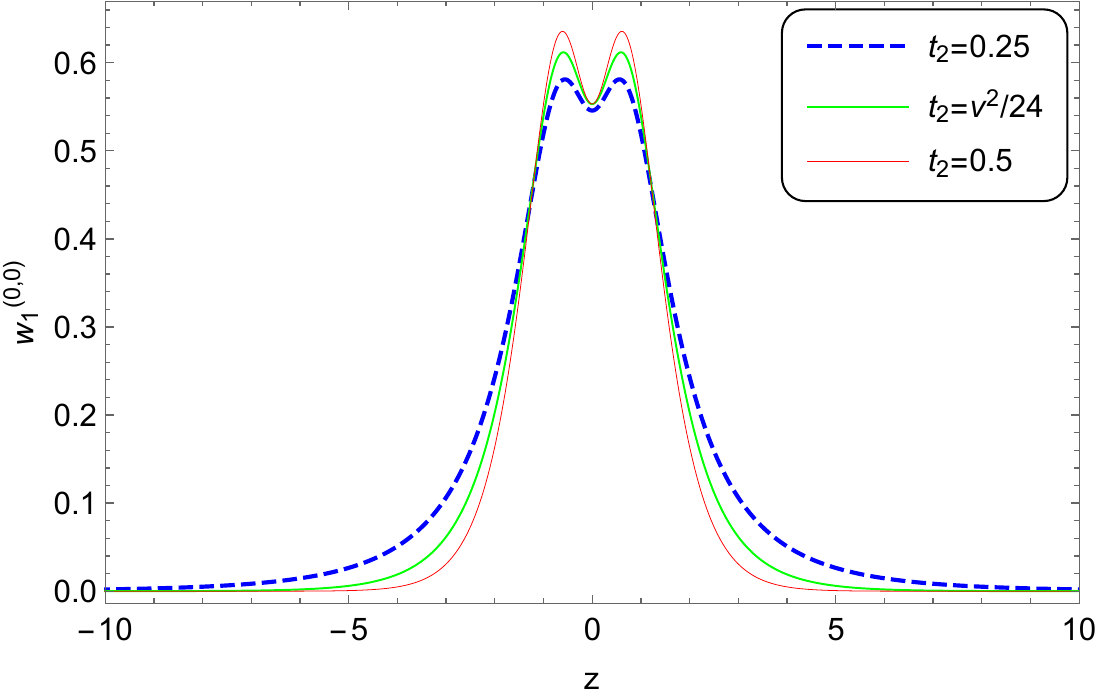}}
\end{center}\vskip -2mm
\caption{The effective potential $U_1(z)$ and the zero mode $w_1^{(0,0)}(z)$ with the parameters
         $k=1$, $v=3$, $t_1=3$ and $t_2=0.25,v^2/24,0.5$.}
 \label{FigEffPotZMVec}
\end{figure}

Furthermore, this equation possesses the zero mode solution
\begin{equation}\label{ZMVec}
  w_1^{(0,0)}(z)=N_1a(z)F(R)^{\frac12}.
\end{equation}
From the 4D effective action (\ref{1actionVec}), it can be seen that the localization of the 4D vector
field requires the integral ${I}_{1}^{(mm')}$ not to be divergent. Based on the decomposition
(\ref{2decompositionVec}), there is
\begin{eqnarray} \label{LocCondVec}
   & &\int d\Theta dzF(R)\left(w_1^{(m,n)}F(R)^{-\frac12}e^{il_n\theta}\right)
                     \left(w_1^{(m,n)}F(R)^{-\frac12}e^{-il_n\theta}\right)      \nonumber    \\
   &=&2\pi R_0\int dz\big|w_1^{(m,n)}\big|^2<\infty.
\end{eqnarray}
So, the localization of the 4D vector field depends on the normalization of the KK modes $w_1^{(m,n)}$
with respect to the coordinate $z$. In terms of the coordinate transformation (\ref{CoordTrans}), the
normalization condition of the zero mode (\ref{ZMVec}) can be expressed as
\begin{eqnarray} \label{NormVec}
    \int|w_1^{(0,0)}|^2dz&=& \int|w_1^{(0,0)}|^2a^{-1}dy                       \nonumber               \\
                         &=& N_1^2\int aF(R)dy=1.
\end{eqnarray}
Substituting the warp factor (\ref{WarpFactor2}) and the coupling function (\ref{FR}) into this condition,
we can get
\begin{eqnarray} \label{ExprNormVec}
  N_1^2\int \text{sech}^{\frac{v^2}{12}}(ky)e^{t_1-t_18^{t_2}
            \left(\frac{(5v^2-12+3(v^2+4)\cosh(4ky))^2\text{sech}^{12}(ky)}{v^4}\right)^{-t_2/2}
            +\frac{v^2}{24}\tanh^2(ky)} dy=1.
\end{eqnarray}
Since the coupling parameters $t_1$ and $t_2$ are both positive, the normalization condition
is automatically satisfied. Consequently, the zero mode $w_1^{(0,0)}(z)$ is localized on the
thick brane. The profiles of this zero mode are shown in fig. \ref{FigZMVec} with specific
values of the parameters. From the figure, it can be seen that the vector zero mode is localized
at the brane position.

For the massive modes, their localization properties are determined by the behavior of the effective
potential (\ref{EffPotVec}). In terms of the coordinate transformation (\ref{CoordTrans}), the
effective potential can be expressed with respect to the coordinate $y$ as
\begin{equation}\label{EffPotVec-y}
  U_1(z(y))=aa''+a'^2+\frac{a^2F''(R)+3aa'F'(R)}{2F(R)}-\frac{a^2F'(R)^2}{4F(R)^2}.
\end{equation}
Then, by substituting eqs. (\ref{WarpFactor2}), (\ref{ScaCurv}) and (\ref{FR}) into the above
expression and performing an asymptotic analysis, we obtain the following asymptotic behaviors
of the effective potential
\begin{eqnarray} \label{AsmpEffPotVec}
U_1(z(y\rightarrow\pm\infty))&\rightarrow&
\left\{
  \begin{array}{ll}
    +\infty, \hspace{0.5cm}   &t_2>v^2/24,  \\
    \text C_1, \hspace{0.5cm}  & t_2=v^2/24,  \\
    0,        &0<t_2<v^2/24
  \end{array}
\right.
\end{eqnarray}
with the positive limit
\begin{eqnarray} \label{LmtPTVec}
  \text C_1=\frac{1}{64}3^{-2-\frac{v^2}{12}}e^{\frac{v^2}{12}}k^2t_1^2v^4\left(\frac{v^2+4}{v^2}\right)^{-\frac{v^2}{12}}.
\end{eqnarray}
The effective potential manifests different asymptotic behaviors, which depend on the value of
the coupling parameter $t_2$. In fig. \ref{FigEffPotVec}, the numerical profiles of the effective
potential are presented with $k=1,v=3,t_1=3$ and $t_2=0.25,v^2/12,0.5$. It can be seen that there
is a critical value $v^2/24$ for the coupling parameter $t_2$. For $0<t_2<v^2/24$, the effective
potential takes the form of a volcano potential. At the critical value $t_2=v^2/24$, it reduces
to a PT potential. For $t_2>v^2/24$, the effective potential becomes an infinitely deep potential
well. Consequently, the localization properties of the massive KK modes depend crucially on the
value of $t_2$. We therefore consider these three cases separately in the following discussion.

\begin{figure}[htb]
\begin{center}
\subfigure[$t_1=30,t_2=0.125$.]{\label{FigResSpecVolVec}
\includegraphics[width= 0.38\textwidth]{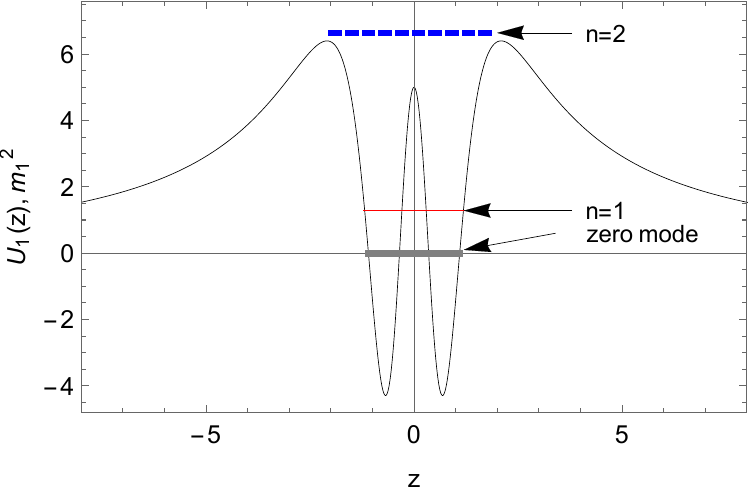}}
\hspace{0.5cm}
\subfigure[$t_1=30,t_2=0.125$.]{\label{FigResPVolVec}
\includegraphics[width= 0.38\textwidth]{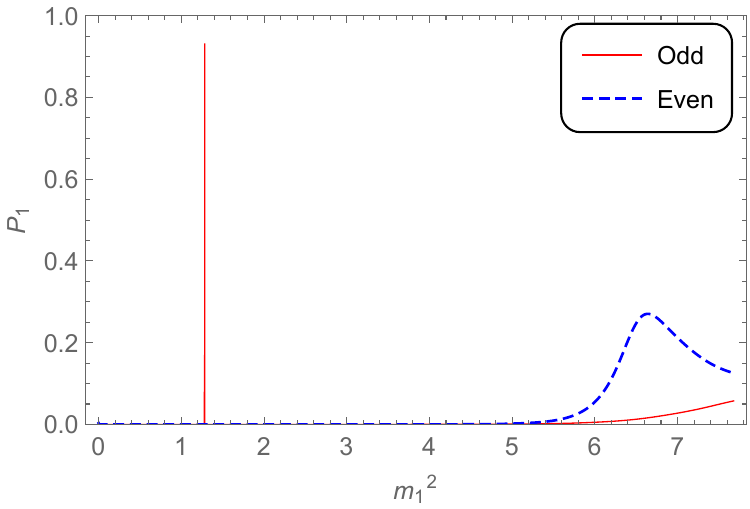}}
\subfigure[$t_1=20,t_2=0.25$.]{\label{2FigResSpecVolVec}
\includegraphics[width= 0.38\textwidth]{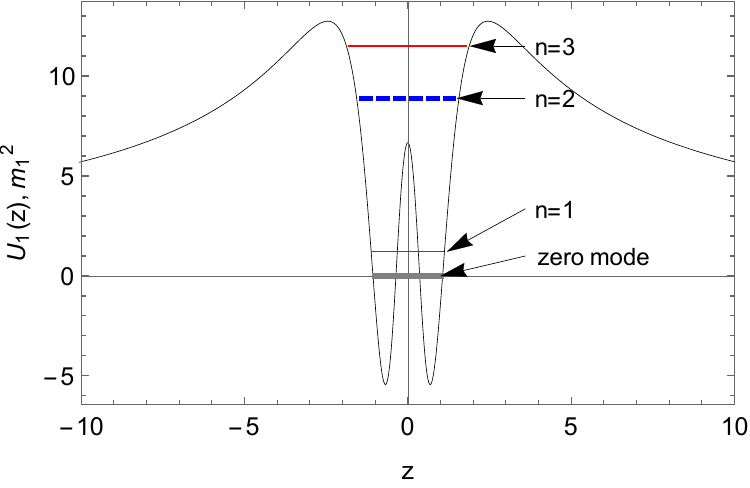}}
\hspace{0.5cm}
\subfigure[$t_1=20,t_2=0.25$.]{\label{2FigResPVolVec}
\includegraphics[width= 0.38\textwidth]{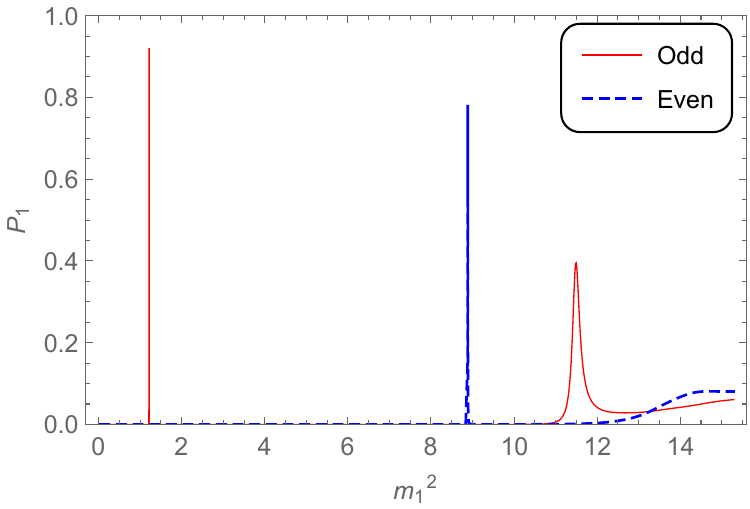}}
\subfigure[$t_1=30,t_2=0.25$.]{\label{3FigResSpecVolVec}
\includegraphics[width= 0.38\textwidth]{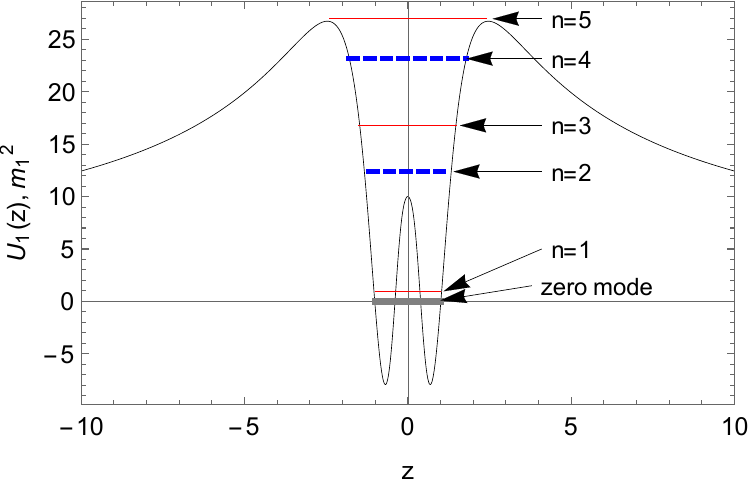}}
\hspace{0.5cm}
\subfigure[$t_1=30,t_2=0.25$.]{\label{3FigResPVolVec}
\includegraphics[width= 0.38\textwidth]{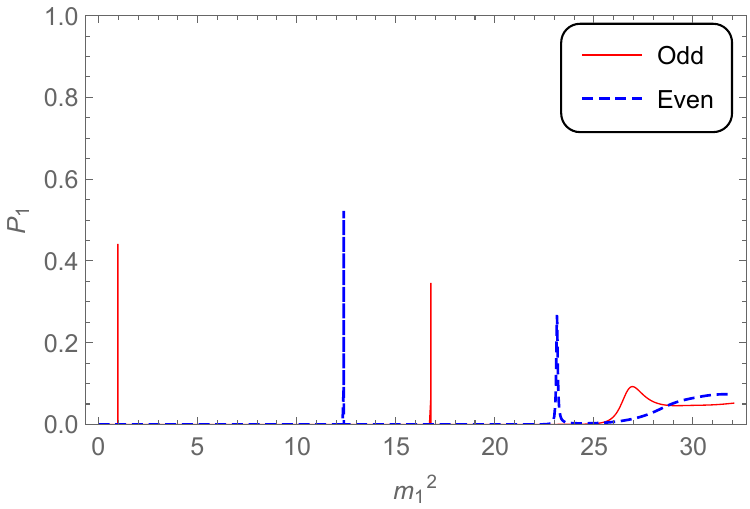}}
\end{center}\vskip -5mm
\caption{The mass spectra, the effective potential $U_1(z)$, and the corresponding relative probability
         $P_1$ with the parameters $k=1$, $v=3$ and $t_1=30,t_2=0.125$; $t_1=20,t_2=0.25$; $t_1=30,t_2=0.25$.
         The effective potential $U_1(z)$ for the black line, the zero mode for the grey line, the even-parity
         resonant KK modes for the blue lines, and the odd-parity resonant KK modes for the red lines.}
 \label{SpecPResoVolVec}
\end{figure}

\begin{figure} %[htb]
\begin{center}
\subfigure[$w_1$.]{\label{Odd1ResVolVec}
\includegraphics[width = 0.31\textwidth]{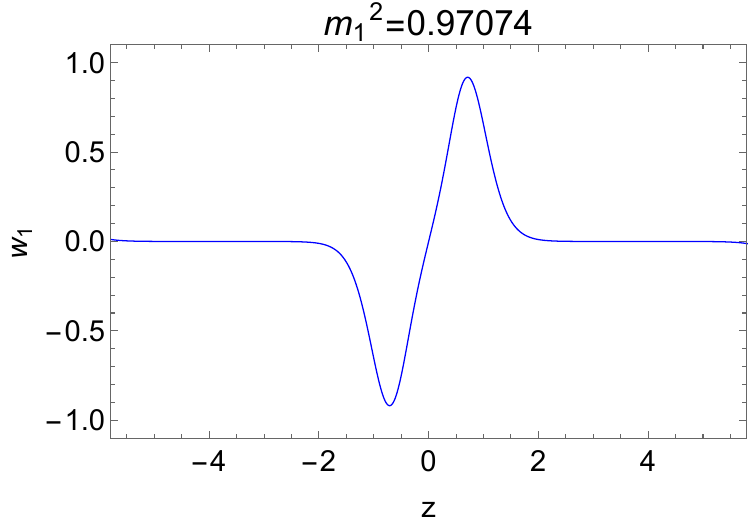}}
\subfigure[$w_1$.]{\label{Even1ResVolVec}
\includegraphics[width = 0.31\textwidth]{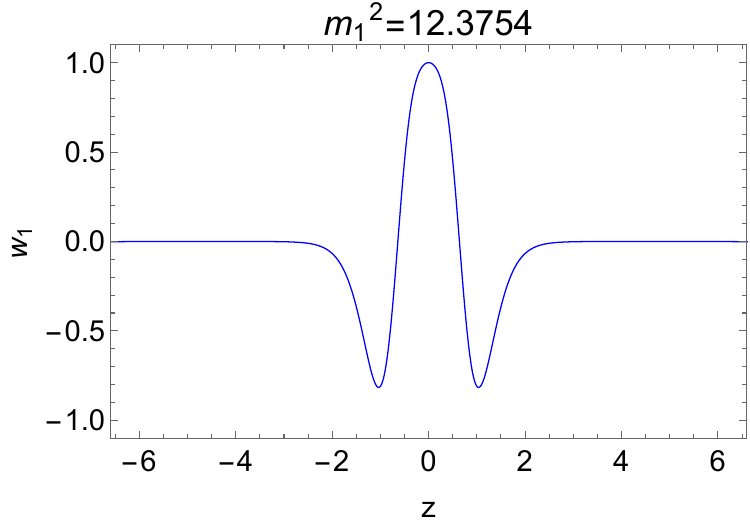}}
\subfigure[$w_1$.]{\label{Odd2ResVolVec}
\includegraphics[width = 0.31\textwidth]{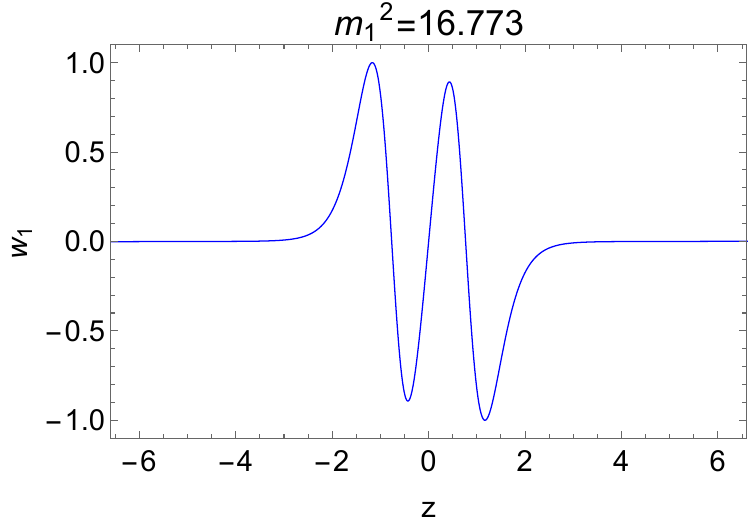}}
\subfigure[$w_1$.]{\label{Even2ResVolVec}
\includegraphics[width = 0.31\textwidth]{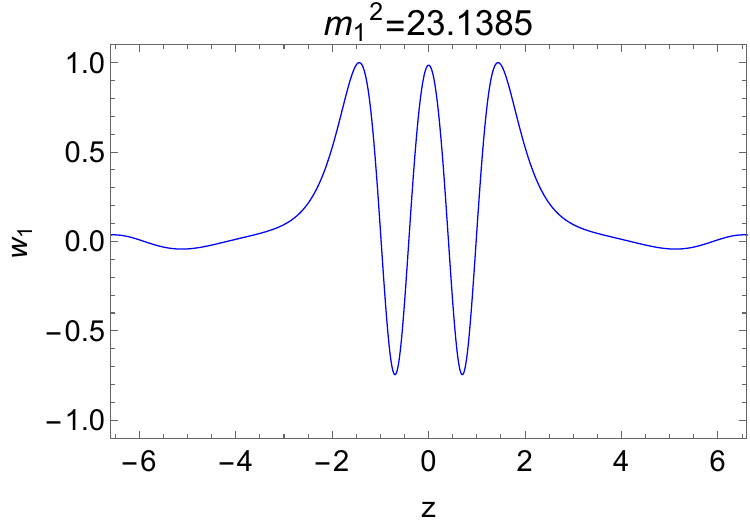}}
\subfigure[$w_1$.]{\label{Odd3ResVolVec}
\includegraphics[width = 0.31\textwidth]{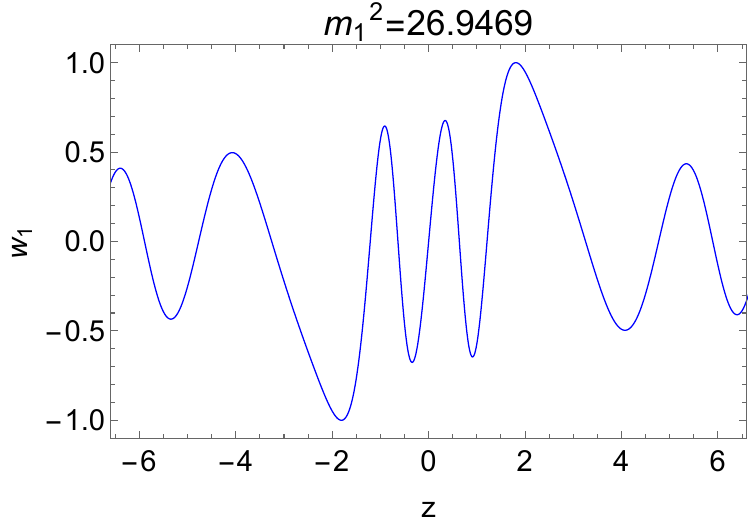}}
\end{center}\vskip -2mm
\caption{The shapes of resonant KK modes $w_1(z)$ for the $U(1)$ gauge vector field
         with different $m_1^{2}$. The parameters are set as $k=1,v=3,
         t_1=30$ and $t_{2}=0.25$.}
 \label{FigResVolVec}
\end{figure}

We first consider the case $0<t_2<v^2/24$, for which the effective potential has a volcano-like
shape. In this scenario, no massive KK modes can be loclaized on the brane. However, the resonant
modes could exist. Using the method described in eq. (\ref{PReso}), we numerically computed the
resonant modes associated with the effective potential (\ref{EffPotVec}) with specific values of
the parameters. The corresponding relative probabilities and mass spectra are displayed in fig. \ref{SpecPResoVolVec}.
In the mass spectra, the ground state corresponds to the localized zero mode, which is a bound
state, and all excited massive states are resonant modes. It can be seen that the number of resonant
modes increases as the coupling parameters $t_1$ and $t_2$ increase. The masses, widths, and lifetimes
of the obtained resonant modes are listed in detail in table \ref{tableRMOriVec}. As an illustration,
the wave functions of the five resonant modes for $t_1=30$ and $t_2=0.25$ are shown in fig. \ref{FigResVolVec}.
%%%%%%%%%%%%%%%%%%%%%%%%%%%%%%%%%%%%%%%%%%%%%%%%%%%%%%%%%%%%%%%%%%%%%%%%%%%%
\begin{table}[tbp]
\centering
\begin{tabular}{|c|c|c|c|c|c|c|c|}
    \hline
    $t_1$               & $t_2$                 & $U^{\text{max}}_{1}$  & $n$         & $m_1^2$ &
    $m_1$               & $\Gamma$              & $\tau$
    \\
    \hline
    $30$                & $0.125$               & $6.3985$              & $1$         & $1.2827$   &
    $1.1326$            & $1.875\times10^{-5}$  & $5.332\times10^{4}$
    \\
                        &                       &                       & $2$         & $6.6406$ &
    $2.5769$            & $0.2445$              & $4.0904$
    \\
    \hline
    $20$                & $0.25$                & $12.7385$             & $1$         & $1.2153$ &
    $1.1024$            & $3.500\times10^{-14}$ & $2.850\times10^{13}$
    \\
    %\cline{3-9}
                        &                       &                       & $2$         & $8.8856$ &
    $2.9809$            & $7.241\times10^{-4}$  & $1.381\times10^3$
    \\
                        &                       &                       & $3$         & $11.4946$ &
    $3.3904$            & $0.0316$              & $31.5977$
    \\
    \hline
    $30$                & $0.25$                & $26.7153$             & $1$         & $0.9707$  &
    $0.9853$            &$8.8983\times10^{-11}$ & $1.124\times10^{10}$
    \\
                        &                       &                       & $2$         & $12.3754$ &
    $3.5179$            & $1.159\times10^{-7}$  & $8.627\times10^6$
    \\
                        &                       &                       & $3$         & $16.7730$ &
    $4.0955$            & $9.160\times10^{-6}$  & $1.092\times10^5$
    \\
                        &                       &                       & $4$         & $23.1385$ &
    $4.8103$            & $9.158\times10^{3}$   & $109.1886$
    \\
                        &                       &                       & $5$         & $26.9469$ &
    $5.1910$            & $0.2400$              & $4.1665$
    \\
    \hline
\end{tabular}
\caption{The mass, width, and lifetime of resonant KK modes $w_1(z)$.
         The parameters are set as $k=1$ and $v=3$. }
    \label{tableRMOriVec}
\end{table}
%****************************************************************

\begin{figure} %[htbp]
\begin{center}
\includegraphics[width= 0.49\textwidth]{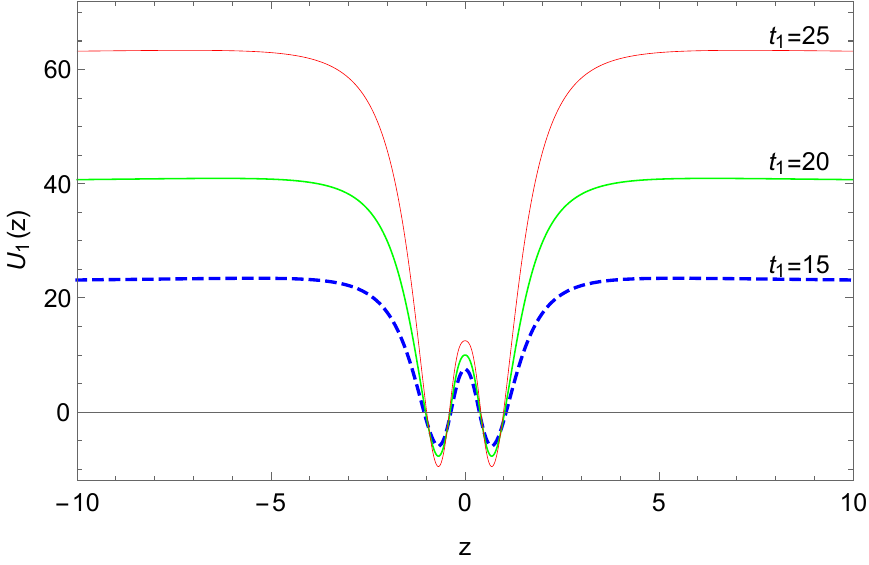}
\caption{The effective potential $U_1(z)$ with the parameter $t_1=15,20$, and $25$.
        The other parameters are set as $k=1,v=3$, and $t_2=3/8$.}
\label{figSpecPTVec}
\end{center}
\end{figure}

We next consider the critical case $t_2=v^2/24$, for which the effective potential takes the form
of a PT potential. This potential tends to the positive constant $ \text C_1$ (\ref{LmtPTVec}) when
far away from the brane. The corresponding effective potentials for certain values of the coupling
parameter $t_1$ are shown in fig. \ref{figSpecPTVec}. It can be seen that the depth of the potential
well increases as $t_1$ rises. For these profiles of the effective potential, the corresponding bound
states were solved numerically, and their mass spectra are given by
\begin{eqnarray}
  m_1^2=&\hspace{-3.8cm}\{0,1.17,9.83,13.35,18.17,20.98\},                                &\text{for}~t_1=15;     \label{SpecPTVec1} \\
  m_1^2=&\hspace{0cm}\{0,0.98,12.30,16.84,23.92,29.25,33.86,37.24,39.49\},\hspace{0.5cm}  &\text{for}~t_1=20;     \label{SpecPTVec2} \\
  m_1^2=&\hspace{-0.7cm}\{0,0.79,14.71,19.93,28.78,36.35,43.27,49.33,54.16,               &\nonumber                                 \\
        &\hspace{-6.7cm}57.94,60.60\},                                                    &\text{for}~t_1=25.     \label{SpecPTVec3}
\end{eqnarray}
Therefore, in the case $t_2=v^2/24$, a finite number of massive KK modes can be localized on the
brane. Moreover, the number of localized massive modes increases with the coupling parameter $t_1$.

\begin{figure} %[htbp]
\begin{center}
\includegraphics[width= 0.49\textwidth]{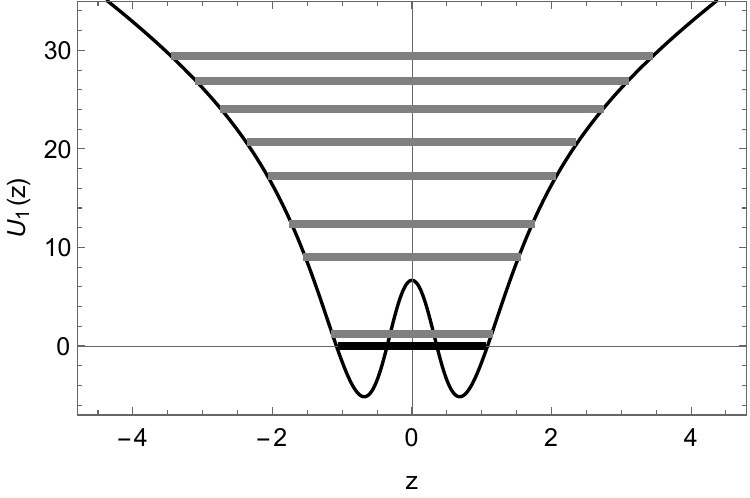}
\caption{The effective potential $U_1(z)$ with the parameters $k=1,v=3,t_1=10$, and $t_2=0.5$.}
\label{figSpecIDWVec}
\end{center}
\end{figure}

Finally, we consider the case $t_2=0.5$, which satisfies $t_2>v^2/24$. In this case, the effective
potential becomes an infinitely deep potential well. Such a potential can localize all KK modes,
including both the massless mode and the massive modes. We numerically solved the low-lying massive
states for specific values of the parameters and present their mass spectrum, together with the
massless mode, in fig. \ref{figSpecIDWVec}. As expected, all massive modes are localized on the
brane and form an infinitely discrete mass spectrum.

Therefore, in the 6D spacetime, the zero mode of the $U(1)$ gauge vector field can be localized on
the brane with codimension-two. The localization properties of the massive KK modes are primarily
determined by the coupling parameter $t_2$, which possesses a critical value $t_2=v^2/24$. For
$0<t_2<v^2/24$, the effective potential has a volcano-like profile. In this case, the massive KK
modes cannot be localized on the brane, but resonant modes may exist, and their number increases
with the coupling parameters $t_1$ and $t_2$. When $t_2=v^2/24$, the effective potential becomes
a PT potential, which supports a finite number of localized massive vector KK modes. The number
of such bound states increases with $t_1$. For $t_2>v^2/24$, the effective potential develops into
an infinitely deep potential well, leading to the localization of all massive KK modes on the thick
brane. Thus, the massive-mode spectrum exhibits qualitatively different localization behaviors in
the three regions of the parameter space separated by the critical value $t_2=v^2/24$.

%%%%%%%%%%%%%%%%%%%%%%%%%%%%%%

\subsection{Kalb-Ramond fields} \label{KR}

In this section, we investigate the localization of the 6D KR field on the brane by taking into
account its coupling to the gravity. The minimal coupling case was studied in ref. \cite{LYT2510.16491},
where it was shown that additional mechanisms are required to localize the KR field on the brane
in a 6D spacetime. Here, we will analyze the localization properties of the KR field in the presence
of the gravity coupling. Besides, in ref. \cite{LYT2510.16491}, it is demonstrated that the gauge
invariance of the 4D effective action of the 6D KR field is preserved through the Stueckelberg
mechanism.

With introducing the gravity coupling, we consider the action of a 6D free KR field $B_{MN}$ as
\begin{equation}\label{6DactionKR}
  S_{\text{KR}}=-\int d^6x\sqrt{-g}F(R)H^{MNL}H_{MNL},
\end{equation}
where the field strength tensor
\begin{equation}\label{fieldStrnthKR}
  H_{MNL}=\partial_MB_{NL}+\partial_LB_{MN}+\partial_NB_{LM}.
\end{equation}
With the KK decomposition
\begin{subequations}\label{decompositionKR}
\begin{eqnarray}
  B_{\mu\nu}    &=& \sum_m\hat B_{\mu\nu}^{(m)}(x^{\sigma}) W_1^{(m)}(z,\Theta)F(R)^{-\frac12},    \\
  B_{\mu z}     &=& \sum_m\hat C_{\mu}^{(m)}(x^{\sigma}) W_2^{(m)}(z,\Theta)F(R)^{-\frac12},       \\
  B_{\mu\Theta} &=& \sum_m\hat D_{\mu}^{(m)}(x^{\sigma}) W_3^{(m)}(z,\Theta)F(R)^{-\frac12},       \\
  B_{z\Theta}   &=& \sum_m\hat{\zeta}^{(m)}(x^{\sigma}) W_4^{(m)}(z,\Theta)F(R)^{-\frac12},
\end{eqnarray}
\end{subequations}
the fundamental 6D action (\ref{6DactionKR}) can be reduced into the following 4D effective one
\begin{eqnarray}\label{4DEffactionKR}
 S_{\text{KR}}&=& -\int d^6x\sqrt{-g}F(R)H^{MNL}H_{MNL}                                              \nonumber   \\
              &=& -\int d^6x\sqrt{-g}F(R)\big(H^{\mu\nu\tau}H_{\mu\nu\tau}+3H^{\mu\nu z}H_{\mu\nu z}
                   +3H^{\mu\nu\Theta}H_{\mu\nu\Theta}+6H^{\mu z\Theta}H_{\mu z\Theta}\big)           \nonumber   \\
              &=& -\sum_m\sum_{m'}\int d^4x\sqrt{-\hat g}\bigg[I_1^{(mm')}\hat H^{\mu\nu\tau(m)}\hat H_{\mu\nu\tau}^{(m')}
                   +\big(I_5^{(mm')}+I_9^{(mm')}\big)\hat B^{\mu\nu(m)}\hat B_{\mu\nu}^{(m')}        \nonumber   \\
              & & +I_2^{(mm')}\hat F^{\mu\nu(m)}\hat F_{\mu\nu}^{(m')}+I_6^{(mm')}\hat G^{\mu\nu(m)}\hat G_{\mu\nu}^{(m')}
                   +I_3^{(mm')}\hat B^{\mu\nu(m)}\hat F_{\mu\nu}^{(m')}                              \nonumber   \\
              & & +I_4^{(mm')}\hat F^{\mu\nu(m)}\hat B_{\mu\nu}^{(m')}+I_7^{(mm')}\hat B^{\mu\nu(m)}\hat G_{\mu\nu}^{(m')}
                   +I_8^{(mm')}\hat G^{\mu\nu(m)}\hat B_{\mu\nu}^{(m')}                              \nonumber   \\
              & & +I_{10}^{(mm')}\partial^{\mu}\hat{\zeta}^{(m)}\partial_{\mu}\hat{\zeta}^{(m')}
                   +I_{11}^{(mm')}\hat C^{\mu(m)}\partial_{\mu}\hat{\zeta}^{(m')}
                   -I_{12}^{(mm')}\hat D^{\mu(m)}\partial_{\mu}\hat{\zeta}^{(m')}                    \nonumber   \\
              & & +I_{13}^{(mm')}\partial^{\mu}\hat{\zeta}^{(m)}\hat C_{\mu}^{(m')}
                   +I_{14}^{(mm')}\hat C^{\mu(m)}\hat C_{\mu}^{(m')}
                   -I_{15}^{(mm')}\hat D^{\mu(m)}\hat C_{\mu}^{(m')}                                 \nonumber   \\
              & & -I_{16}^{(mm')}\partial^{\mu}\hat{\zeta}^{(m)}\hat D_{\mu}^{(m')}
                   -I_{17}^{(mm')}\hat C^{\mu(m)}\hat D_{\mu}^{(m')}
                   +I_{18}^{(mm')}\hat D^{\mu(m)}\hat D_{\mu}^{(m')}\bigg],
\end{eqnarray}
where the field strengths
\begin{eqnarray}
  \hat H_{\mu\nu\tau}^{(m)}&=& \partial_{\mu}\hat B_{\nu\tau}^{(m)}+\partial_{\tau}\hat B_{\mu\nu}^{(m)}
                                 +\partial_{\nu}\hat B_{\tau\mu}^{(m)},                       \label{tensor1KR}     \\
  \hat F_{\mu\nu}^{(m)}&=& \partial_{\mu}\hat C_{\nu}^{(m)}-\partial_{\nu}\hat C_{\mu}^{(m)}, \label{tensor2KR}     \\
  \hat G_{\mu\nu}^{(m)}&=& \partial_{\mu}\hat D_{\nu}^{(m)}-\partial_{\nu}\hat D_{\mu}^{(m)}, \label{tensor3KR}
\end{eqnarray}
and the constants are given by
\begin{subequations}\label{constantKR}
\begin{eqnarray}
&&{I}_{1}^{(mm')}\equiv\int d \Theta d z~ W_{1}^{(m)} W_{1}^{(m')} ,                 \\
&&{I}_{2}^{(mm')}\equiv3 \int {d} \Theta {d} z~  W_{2}^{(m)} W_{2}^{(m')},           \\
&&{I}_{3}^{(mm')}\equiv3 \int {d} \Theta {d} z~ \left(\partial_zW_{1}^{(m)}\right) {W}_{2}^{(m')} ,         \\
&&{I}_{4}^{(mm')}\equiv3 \int {d} \Theta {d} z~  {W}_{2}^{(m)}\left(\partial_zW_{1}^{(m')}\right),                   \\
&&{I}_{5}^{(mm')}\equiv3 \int {d\Theta} {d} z~ \left(\partial_zW_{1}^{(m)}\right)
                          \left(\partial_zW_{1}^{(m')}\right),                 \\
&&{I}_{6}^{(mm')}\equiv3 \int d \Theta d z~ W_{3}^{(m)} W_{3}^{(m')} ,     \\
&&{I}_{7}^{(mm')}\equiv3 \int {d} \Theta {d} z~ \left(\partial_{\Theta}{W}_{1}^{(m)}\right)
        {W}_{3}^{(m')},                                                        \\
&&{I}_{8}^{(mm')}\equiv3 \int {d\Theta dz}~ {W}_{3}^{(m)}\left( \partial_{\Theta}{W}_{1}^{(m')} \right),\\
&&{I}_{9}^{(mm')}\equiv3 \int d \Theta d z~ \left(\partial_{\Theta}W_{1}^{(m)}\right)
                          \left(\partial_{\Theta}W_{1}^{(m')}\right),                    \\
&&{I}_{10}^{(mm')}\equiv6 \int {d} \Theta {d} z~ {W}_{4}^{(m)} {W}_{4}^{(m')},             \\
&&{I}_{11}^{(mm')}\equiv6 \int {d\Theta} {d} z~ \left(\partial_{\Theta}{W}_{2}^{(m)}\right)
                           {W}_{4}^{(m')},   \\
&&{I}_{12}^{(mm')}\equiv6 \int {d\Theta} {d} z~ \left(\partial_z{W}_{3}^{(m)}\right)
                            {W}_{4}^{(m')},   \\
&&{I}_{13}^{(mm')}\equiv6 \int {d\Theta} {d} z~ {W}_{4}^{(m)}
                           \left(\partial_{\Theta}{W}_{2}^{(m')}\right) ,           \\
&&{I}_{14}^{(mm')}\equiv6 \int {d\Theta} {d} z~ \left(\partial_{\Theta}{W}_{2}^{(m)}\right)
                           \left(\partial_{\Theta}{W}_{2}^{(m')}\right),            \\
&&{I}_{15}^{(mm')}\equiv6 \int {d\Theta} {d} z~ \left(\partial_z{W}_{3}^{(m)}\right)
                           \left(\partial_{\Theta}{W}_{2}^{(m')}\right),            \\
&&{I}_{16}^{(mm')}\equiv2 \int {d\Theta} {d} z~ {W}_{4}^{(m)}
                           \left(\partial_z{W}_{3}^{(m')}\right),                   \\
&&{I}_{17}^{(mm')}\equiv6 \int {d\Theta} {d} z~ \left(\partial_{\Theta}{W}_{2}^{(m)}\right)
                           \left(\partial_z{W}_{3}^{(m')}\right),                   \\
&&{I}_{18}^{(mm')}\equiv6 \int {d\Theta} {d} z~ \left(\partial_z{W}_{3}^{(m)}\right)
                           \left(\partial_z{W}_{3}^{(m')}\right).
\end{eqnarray}
\end{subequations}
In the KK decomposition (\ref{decompositionKR}), $\hat B_{\mu\nu}^{(m)}(x^{\sigma})$ denotes the
4D KR field, $\hat C_{\mu}^{(m)}(x^{\sigma})$ and $\hat D_{\mu}^{(m)}(x^{\sigma})$ represent two
4D vector fields, and $\hat{\zeta}^{(m)}(x^{\sigma})$ is a 4D scalar field. In this effective
action (\ref{4DEffactionKR}), the combination $I_5^{(mm')}+I_9^{(mm')}$ contributes to the mass
term of the 4D KR field.

By varying the 4D effective action (\ref{4DEffactionKR}) with respect to $\hat B_{\mu\nu}^{(m)}$,
$\hat C_{\mu}^{(m)}$, $\hat D_{\mu}^{(m)}$ and $\hat{\zeta}^{(m)}$, we can obtain
\begin{subequations}\label{vary4DactionKR}
\begin{align}
  \frac{I_1^{(mm')}}{\sqrt{-\hat g}}\partial_{\lambda}\hspace{-0.02cm}\left(\sqrt{-\hat g}\hat H^{\mu\nu\lambda(m)}\right) \hspace{-0.05cm}
     -\hspace{-0.05cm}\left(I_5^{(mm')}\hspace{-0.05cm}+I_9^{(mm')}\right)\hat B^{\mu\nu(m)}\hspace{-0.05cm}
     -I_4^{(mm')}\hat F^{\mu\nu(m)}\hspace{-0.05cm}                         %\hspace{-0.1cm}         \nonumber   \\
     -\hspace{-0.05cm}I_8^{(mm')}\hat G^{\mu\nu(m)}=0,                                                                       \\
  \frac{I_2^{(mm')}}{\sqrt{-\hat g}}\partial_{\nu}\left(\sqrt{-\hat g}\hat F^{\mu\nu(m)}\right)
     +\frac{I_3^{(mm')}}{\sqrt{-\hat g}}\partial_{\nu}\left(\sqrt{-\hat g}\hat B^{\mu\nu(m)}\right)
     +I_{13}^{(mm')}\partial^{\mu}\hat{\zeta}^{(m)}+I_{14}^{(mm')}\hat C_{\mu}^{(m)}         \hspace{1cm}         \nonumber   \\
     -I_{15}^{(mm')}\hat D_{\mu}^{(m)}=0,                                   \\
  \frac{I_6^{(mm')}}{\sqrt{-\hat g}}\partial_{\nu}\left(\sqrt{-\hat g}\hat G^{\mu\nu(m)}\right)
     +\frac{I_7^{(mm')}}{\sqrt{-\hat g}}\partial_{\nu}\left(\sqrt{-\hat g}\hat B^{\mu\nu(m)}\right)
     -I_{16}^{(mm')}\partial^{\mu}\hat{\zeta}^{(m)}-I_{17}^{(mm')}\hat C_{\mu}^{(m)}         \hspace{1cm}         \nonumber    \\
     +I_{18}^{(mm')}\hat D_{\mu}^{(m)}=0,                                   \\
  I_{10}^{(mm')}\partial_{\mu}\left(\sqrt{-\hat g}\partial^{\mu}\hat{\zeta}^{(m)}\right)
     +I_{11}^{(mm')}\partial_{\mu}\left(\sqrt{-\hat g}\hat C_{\mu}^{(m)}\right)
     -I_{12}^{(mm')}\partial_{\mu}\left(\sqrt{-\hat g}\hat D_{\mu}^{(m)}\right)=0.
\end{align}
\end{subequations}
On the other hand, by varying the 6D action (\ref{6DactionKR}) with respect to the 6D KR field $B_{MN}$,
we can obtain the equation of motion
\begin{equation} \label{EoMKR}
  \frac{1}{\sqrt{-g}}\partial_M\left(\sqrt{-g}F(R)H^{MNL}\right)=0.
\end{equation}
This equation possesses the following four component equations
\begin{subequations} \label{EoMcomponentKR}
\begin{align}
  &\frac{1}{\sqrt{-\hat g}}\partial_{\lambda}\left(\sqrt{-\hat g}\hat H^{\mu\nu\lambda(m)}\right)
    +(\lambda_1+\lambda_2)\hat B^{\mu\nu(m)}+\lambda_3\hat F^{\mu\nu(m)}+\lambda_4\hat G^{\mu\nu(m)}=0,        \\
  &\frac{1}{\sqrt{-\hat g}}\partial_{\nu}\left(\sqrt{-\hat g}\hat F^{\mu\nu(m)}\right)
    +\frac{\lambda_5}{\sqrt{-\hat g}}\partial_{\nu}\left(\sqrt{-\hat g}\hat B^{\mu\nu(m)}\right)
    -\lambda_6\partial^{\mu}\hat{\zeta}^{(m)}-\lambda_7\hat C^{\mu(m)}+\lambda_8\hat D^{\mu(m)}=0,             \\
  &\frac{1}{\sqrt{-\hat g}}\partial_{\nu}\left(\sqrt{-\hat g}\hat G^{\mu\nu(m)}\right)
    +\frac{\lambda_9}{\sqrt{-\hat g}}\partial_{\nu}\left(\sqrt{-\hat g}\hat B^{\mu\nu(m)}\right)  \hspace{-0.05cm}
    +\lambda_{10}\partial^{\mu}\hat{\zeta}^{(m)}+\lambda_{11}\hat C^{\mu(m)}                     \hspace{-0.05cm}
    -\lambda_{12}\hat D^{\mu(m)}=0,                                                                            \\
  &\partial_{\mu}\left(\sqrt{-g}\partial^{\mu}\hat{\zeta}^{(m)}\right)
    +\lambda_{13}\partial_{\mu}\left(\sqrt{-g}\hat C^{\mu(m)}\right)
    -\lambda_{14}\partial_{\mu}\left(\sqrt{-g}\hat D^{\mu(m)}\right)=0,
\end{align}
\end{subequations}
where
\begin{eqnarray} \label{lambdaF}
  &&\lambda_1 \equiv\frac{\partial_z\left(F(R)\partial_z\left(W_1^{(m)}F(R)^{-\frac12}\right)\right)}
                    {W_1^{(m)}F(R)^{\frac12}},                               \hspace{0.2cm}
    \lambda_2 \equiv\frac{\partial^2_{\Theta}W_1^{(m)}}{W_1^{(m)}},                               \nonumber      \\
  &&\lambda_3 \equiv\frac{\partial_z\left(W_2^{(m)}F(R)^{\frac12}\right)}
                    {W_1^{(m)}F(R)^{\frac12}},                              \hspace{2.2cm}
    \lambda_4 \equiv\frac{\partial_{\Theta}W_3^{(m)}}{W_1^{(m)}},                       \nonumber      \\
  &&\lambda_5 \equiv\frac{\partial_z\left(W_1^{(m)}F(R)^{-\frac12}\right)}
                    {W_2^{(m)}F(R)^{-\frac12}},                              \hspace{1.95cm}
    \lambda_6 \equiv\frac{\partial_{\Theta}W_4^{(m)}}{W_2^{(m)}},                       \nonumber      \\
  &&\lambda_7 \equiv\frac{\partial^2_{\Theta}W_2^{(m)}}{W_2^{(m)}},            \hspace{3.75cm}
    \lambda_8 \equiv\frac{\partial_{\Theta}\partial_z\left(W_3^{(m)}F(R)^{-\frac12}\right)}
                    {W_2^{(m)}F(R)^{-\frac12}},                              \nonumber      \\
  &&\lambda_9 \equiv\frac{\partial_{\Theta}W_1^{(m)}}{W_3^{(m)}},              \hspace{3.65cm}
    \lambda_{10}\equiv\frac{\partial_z\left(W_4^{(m)}F(R)^{\frac12}\right)}
                    {W_3^{(m)}F(R)^{\frac12}},                      \nonumber      \\
  &&\lambda_{11}\equiv\frac{\partial_z\left((\partial_{\Theta}W_2^{(m)})F(R)^{\frac12}\right)}
                    {W_3^{(m)}F(R)^{\frac12}},                         \hspace{1.2cm}
    \lambda_{12}\equiv\frac{\partial_z\left(F(R)\partial_z\left(W_3^{(m)}F(R)^{-\frac12}\right)\right)}
                    {W_3^{(m)}F(R)^{\frac12}},                     \nonumber      \\
  &&\lambda_{13}\equiv\frac{\partial_{\Theta}W_2^{(m)}}{W_4^{(m)}},    \hspace{3.55cm}
    \lambda_{14}\equiv\frac{\partial_z\left(W_3^{(m)}F(R)^{-\frac12}\right)}{W_4^{(m)}F(R)^{-\frac12}}.
\end{eqnarray}
Equations (\ref{vary4DactionKR}) and (\ref{EoMcomponentKR}) are both derived from the 6D
action (\ref{6DactionKR}). Therefore, these two sets of equations should be mutually compatible,
which leads to the following conditions:
\begin{eqnarray} \label{ConsisRelasn}
  &&I_1^{(mm')}=\delta^{mm'},                    \hspace{1.2cm}
    I_2^{(mm')}=\delta^{mm'},                    \hspace{1.2cm}
    I_3^{(mm')}=\lambda_5\delta^{mm'},           \hspace{0.8cm}
    I_4^{(mm')}=-\lambda_3\delta^{mm'},                              \nonumber    \\
  &&I_5^{(mm')}=-\lambda_1\delta^{mm'},          \hspace{0.5cm}
    I_6^{(mm')}=\delta^{mm'},                    \hspace{1.2cm}
    I_7^{(mm')}=\lambda_9\delta^{mm'},           \hspace{0.8cm}
    I_8^{(mm')}=-\lambda_4\delta^{mm'},                              \nonumber    \\
  &&I_9^{(mm')}=-\lambda_2\delta^{mm'},          \hspace{0.5cm}
    I_{10}^{(mm')}=\delta^{mm'},                 \hspace{1.2cm}
    I_{11}^{(mm')}=\lambda_{13}\delta^{mm'},     \hspace{0.6cm}
    I_{12}^{(mm')}=\lambda_{14}\delta^{mm'},                         \nonumber    \\
  &&I_{13}^{(mm')}=-\lambda_{6}\delta^{mm'},     \hspace{0.5cm}
    I_{14}^{(mm')}=-\lambda_{7}\delta^{mm'},     \hspace{0.5cm}
    I_{15}^{(mm')}=-\lambda_{8}\delta^{mm'},     \hspace{0.5cm}
    I_{16}^{(mm')}=-\lambda_{10}\delta^{mm'},                        \nonumber    \\
  &&I_{17}^{(mm')}=-\lambda_{11}\delta^{mm'},    \hspace{0.39cm}
    I_{18}^{(mm')}=-\lambda_{12}\delta^{mm'}.
\end{eqnarray}
These conditions indicate that the matrices in eq. (\ref{constantKR}) are diagonal and incorporate
the assumption that the 4D KR field is localized on the thick brane. Therefore, they impose constraints
on the higher-dimensional model, ensuring that the resulting higher-dimensional theory remains
consistent with observational requirements.

By canonically normalizing the action (\ref{4DEffactionKR}), the 4D effective action can be written
as
\begin{equation}\label{2actionKR}
S_{\text{KR}}=-\sum_m\sum_{m'}\int {d^{4}x} \sqrt{-\hat{g}}
     \;\bigg(
            \;\hat{H}^{\mu\nu\tau(m)}\;\hat{H}_{\mu\nu\tau}^{(m')}
            +\frac{I_{5}^{(mm')}+I_{9}^{(mm')}}{I_{1}^{(mm')}}\;\hat{B}^{\mu\nu(m)}\hat{B}_{\mu\nu}^{(m')}+\cdots\bigg).
\end{equation}
Taking the consistency conditions (\ref{ConsisRelasn}) into this canonical action, we have
\begin{equation}\label{CorresKR}
  \lambda_1+\lambda_2=-\frac{I_{5}^{(mm')}+I_{9}^{(mm')}}{I_{1}^{(mm')}},
\end{equation}
and the effective mass $m$ is given by
\begin{equation}\label{massKR}
  m^2=-(\lambda_1+\lambda_2).
\end{equation}

We further separate variables as
\begin{eqnarray}\label{2decompositionKR}
W_1^{(m)}(z,\Theta)&=&\sum_{n}u^{(m,n)}_1(z)e^{il_n\Theta},\\
W_2^{(m)}(z,\Theta)&=&\sum_{n}u^{(m,n)}_2(z)e^{il_n\Theta},\\
W_3^{(m)}(z,\Theta)&=&\sum_{n}u^{(m,n)}_3(z)e^{il_n\Theta},\\
W_4^{(m)}(z,\Theta)&=&\sum_{n}u^{(m,n)}_4(z)e^{il_n\Theta},
\end{eqnarray}
where $u^{(m,n)}_j(z),j=1,2,3,4$ are the KK modes of the four 4D fields with respect to the large
extra dimension $z$. Then, eqs. (\ref{lambdaF})
can be rewritten as
\begin{eqnarray} \label{lambdauF}
  \lambda_1u_1 &\equiv& \frac{\partial_z\left(F(R)\partial_z\left(u_1F(R)^{-\frac12}\right)\right)}
                    {F(R)^{\frac12}},               \hspace{0.78cm}
    \lambda_2 \equiv -l_n^2,                                                        \nonumber      \\
  \lambda_3u_1 &\equiv& \frac{\partial_z\left(u_2F(R)^{\frac12}\right)}{F(R)^{\frac12}},   \hspace{2.4cm}
    \lambda_4u_1 \equiv il_nu_3,                                                    \nonumber      \\
  \lambda_5u_2 &\equiv& \partial_z\left(u_1F(R)^{-\frac12}\right)F(R)^{\frac12},   \hspace{1.08cm}
    \lambda_6u_2 \equiv il_nu_4,            \nonumber      \\
  \lambda_7 &\equiv& -l_n^2,   \hspace{4.3cm}
    \lambda_8u_2 \equiv il_n\partial_z\left(u_3F(R)^{-\frac12}\right)F(R)^{\frac12},               \nonumber      \\
  \lambda_9u_3 &\equiv& il_n u_1,            \hspace{3.92cm}
    \lambda_{10}u_3 \equiv \frac{\partial_z\left(u_4F(R)^{\frac12}\right)}{F(R)^{\frac12}},                \nonumber      \\
  \lambda_{11}u_3 &\equiv& \frac{il_n\partial_z\left(u_2F(R)^{\frac12}\right)}{F(R)^{\frac12}},   \hspace{1.8cm}
    \lambda_{12}u_3 \equiv \frac{\partial_z\left(F(R)\partial_z\left(u_3F(R)^{-\frac12}\right)\right)}
                    {F(R)^{\frac12}},                                                  \nonumber      \\
  \lambda_{13}u_4 &\equiv& il_nu_2,   \hspace{3.93cm}
    \lambda_{14}u_4 \equiv \partial_z\left(u_3F(R)^{-\frac12}\right)F(R)^{\frac12}.
\end{eqnarray}
%\nonumber      \\
%  \lambda_7 \equiv -l_n^2,                                                            \hspace{0cm}
%    \lambda_8u_2 \equiv il_n\partial_z\left(u_3(F(R))^{-\frac12}\right)(F(R))^{\frac12}},     \nonumber      \\
%  \lambda_9u_3 \equiv il_n u_1,                                                        \hspace{0cm}
%    \lambda_{10}u_3 \equiv partial_z\left(u_4(F(R))^{\frac12}\right)(F(R))^{-\frac12},\nonumber      \\
%  \lambda_{11}u_3 \equiv il_n\partial_z\left(u_2(F(R))^{\frac12}\right)F(R))^{-\frac12},\hspace{0cm}
%    \lambda_{12}u_3 \equiv \partial_z\left(F(R)\partial_z\left(u_3(F(R))^{-\frac12}\right)\right)
%                    (F(R))^{-\frac12},                                               \nonumber      \\
%  \lambda_{13}u_4 \equiv il_nu_2,                                                       \hspace{0cm}
%    \lambda_{14}u_4 \equiv \partial_z\left(u_3^{(m)}(F(R))^{-\frac12}\right)(F(R))^{\frac12}.

Substituting $\lambda_1$ and $\lambda_2$ defined in eq. (\ref{lambdauF}) into
eq. (\ref{massKR}), we can get
\begin{equation}\label{SchroKR}
\left[ {-\partial_z^2 + U_2(z)} \right]u_1^{(m,n)} =\big(m^2-{l_n^2}\big)u_1^{(m,n)}~,
\end{equation}
where the effective potential
\begin{equation}\label{EffPotKR}
  U_2(z)=\frac{F''(R)}{2F(R)}-\frac{F'(R)^2}{4F(R)^2}.
\end{equation}
The Schr\"{o}dinger-like equation (\ref{SchroKR}) can be recast to
\begin{equation}\label{FactSchroKR}
  \left(\partial_z + \Gamma'(z)\right)\left(-\partial_z + \Gamma'(z)\right)u_1^{(m,n)}
             =m_2^2u_1^{(m,n)}
\end{equation}
with
\begin{eqnarray}\label{Gammam2}
   \Gamma'(z)  =  \frac{F'(R)}{2F(R)},
\end{eqnarray}
where we have defined $m_2^2 = m^2-{l_n^2}$ for convenience. This equation can be written in the
form of supersymmetric form $B^{\dagger}Bu_1^{(m,n)} =m_2^2u_1^{(m,n)}$ with $B= -\partial_z +\Gamma'(z)$.
Therefore, the eigenvalues satisfy $m_2^2\geq0$, and no tachyonic unstable modes exist.

\begin{figure} %[htb]
\begin{center}
\subfigure[$U_2(z).$] {\label{FigEffPotKR}
\includegraphics[width= 0.45\textwidth]{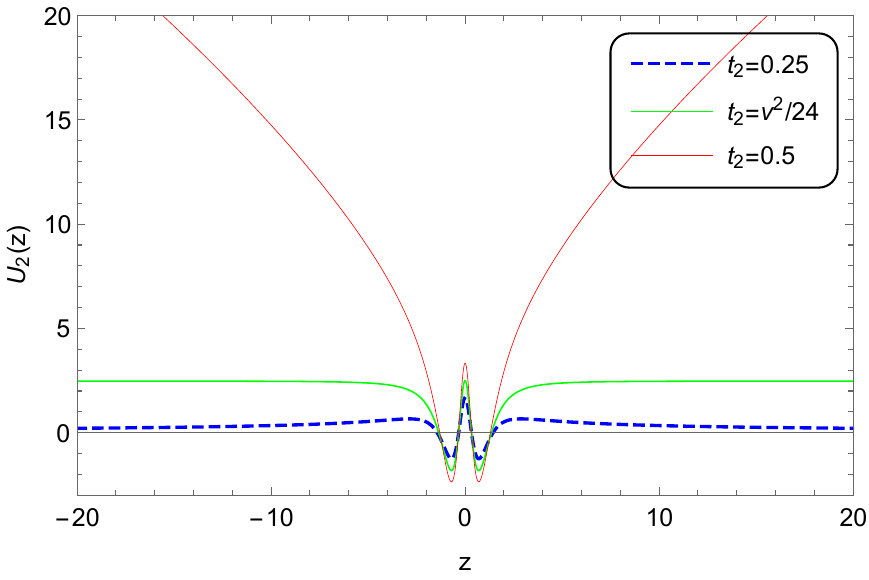}}
\subfigure[$u_1^{(0,0)}(z).$] {\label{FigZMKR}
\includegraphics[width= 0.45\textwidth]{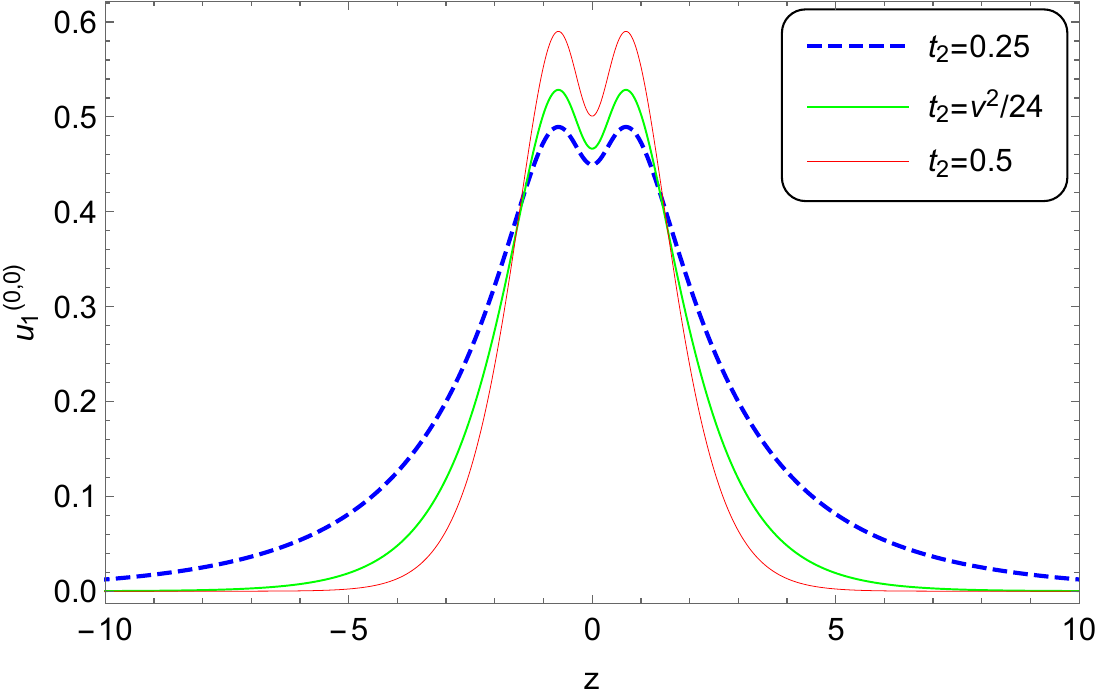}}
\end{center}\vskip -2mm
\caption{The effective potential $U_2(z)$ and the zero mode $u_1^{(0,0)}(z)$ with parameters
         $k=1$, $v=3$, $t_1=3$ and $t_2=0.25,v^2/24,0.5$.}
 \label{FigEffPotZMKR}
\end{figure}

Furthermore, this equation (\ref{FactSchroKR}) can give rise to the zero mode solution
\begin{equation}\label{ZMKR}
  u_1^{(0,0)}(z)=N_2F(R)^{\frac12},
\end{equation}
where $N_2$ is the normalized constant. From the 4D effective action (\ref{4DEffactionKR}), the
localization of the 4D KR field requires the integral $I_1^{(mm')}<\infty$. In light of the decomposition
(\ref{2decompositionKR}), we can get
\begin{eqnarray}\label{LocCondNormKR}
  \int d\Theta dz \left(u_1^{(m,n)}e^{il_n\Theta}\right)
       \left(u_1^{(m,n)}e^{-il_n\Theta}\right)=2\pi R_0\int dz \big|u_1^{(m,n)}\big|^2<\infty.
\end{eqnarray}
This expression imply that the localization of the 4D KR field depends on the normalization of the
KK mode $u_1^{(m,n)}$ with respect to the coordinate $z$. Based on the coordinate transformation
(\ref{CoordTrans}), the normalization condition of the zero mode (\ref{ZMKR}) is
\begin{eqnarray}\label{NormKR}
  \int |u_1^{(0,0)}|^2dz&=& \int |u_1^{(0,0)}|^2a^{-1}dy              \nonumber        \\
                        &=& N_2^2\int F(R)a^{-1}dy=1.
\end{eqnarray}
From this condition, it can be seen that in the minimal coupling case with $F(R)\equiv1$, the
normalization condition cannot be satisfied, and the KR field cannot be localized on the thick
brane. In contrast, when the coupling between the KR field and the gravity is taken into account,
substituting eqs. (\ref{WarpFactor2}), (\ref{ScaCurv}) and (\ref{FR}) into the normalization
condition (\ref{NormKR}) yields
\begin{eqnarray}\label{NormCondKR}
  N_2^2\int \text{sech}^{-\frac{v^2}{12}}(ky)e^{t_1-t_18^{t_2}
            \left(\frac{(5v^2-12+3(v^2+4)\cosh(4ky))^2\text{sech}^{12}(ky)}{v^4}\right)^{-t_2/2}
            -\frac{v^2}{24}\tanh^2(ky)} dy=1.   \hspace{0.2cm}
\end{eqnarray}
From this expression, one finds that the normalization condition can be satisfied for positive values
of the coupling parameters $t_1$ and $t_2$. As a result, the zero mode $u_1^{(0,0)}(z)$ (\ref{ZMKR})
is normalizable. The profiles of this zero mode with specific values of the parameters are presented
in fig. \ref{FigZMKR}. Therefore, the zero mode of the KR field can be localized on the thick brane.

To investigate the localization properties of the massive modes of the KR field, we analyze the
behavior of the effective potential (\ref{EffPotKR}). Based on the coordinate transformation (\ref{CoordTrans}),
the effective potential can be expressed in terms of the coordinate $y$ as
\begin{equation}\label{EffPotKR-y}
  U_2(z(y))=\frac{a^2F''(R)+aa'F'(R)}{2F(R)}-\frac{a^2F'(R)^2}{4F(R)^2}.
\end{equation}
Substituting eqs. (\ref{WarpFactor2}), (\ref{ScaCurv}) and (\ref{FR}) into this potential, we can
further obtain its asymptotic behaviors
\begin{eqnarray}\label{AsymEffPotKR}
U_2(z(y\rightarrow\pm\infty))&\rightarrow&
\left\{
  \begin{array}{ll}
    +\infty, \hspace{0.5cm}   &t_2>v^2/24,  \\
    \text C_2, \hspace{0.5cm}  & t_2=v^2/24,  \\
    0,        &0<t_2<v^2/24
  \end{array}
\right.
\end{eqnarray}
with the positive limit
\begin{eqnarray} \label{LmtPTKR}
  \text C_2=\frac{1}{64}3^{-2-\frac{v^2}{12}}e^{\frac{v^2}{12}}k^2t_1^2v^4\left(\frac{v^2+4}{v^2}\right)^{-\frac{v^2}{12}}.
\end{eqnarray}
From this expression, it can be seen that the asymptotic behavior of the effective potential when
far away from the brane depends on the value of the coupling parameter $t_2$, leading to three
distinct cases. The corresponding potential profiles are plotted numerically in fig. \ref{FigEffPotKR}
for the parameter values $k=1$, $v=3$, $t_1=3$ and $t_2=0.25,v^2/24,0.5$. As shown in the figure,
there exists a critical value $t_2=v^2/24$. For $0<t_2<v^2/24$, the effective potential tends to
zero as $z\rightarrow\pm\infty$, exhibiting a volcano-like structure. When $t_2=v^2/24$, the
effective potential takes the form of a PT potential. For $t_2>v^2/24$, the effective potential
becomes an infinitely deep potential well. Therefore, the localization properties of the massive
KK modes should be analyzed separately for these three cases.

\begin{figure}[htb]
\begin{center}
\subfigure[$t_1=40,t_2=0.125$.]{\label{FigResSpecVolKR}
\includegraphics[width= 0.38\textwidth]{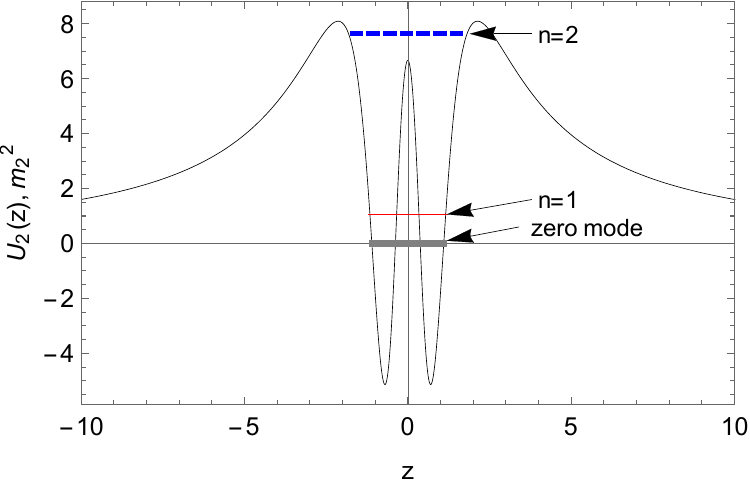}}
\hspace{0.5cm}
\subfigure[$t_1=40,t_2=0.125$.]{\label{FigResPVolKR}
\includegraphics[width= 0.38\textwidth]{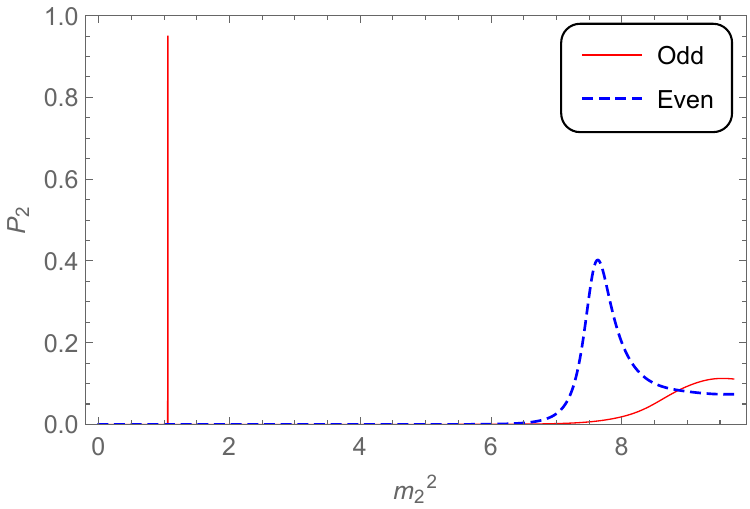}}
\subfigure[$t_1=20,t_2=0.25$.]{\label{2FigResSpecVolKR}
\includegraphics[width= 0.38\textwidth]{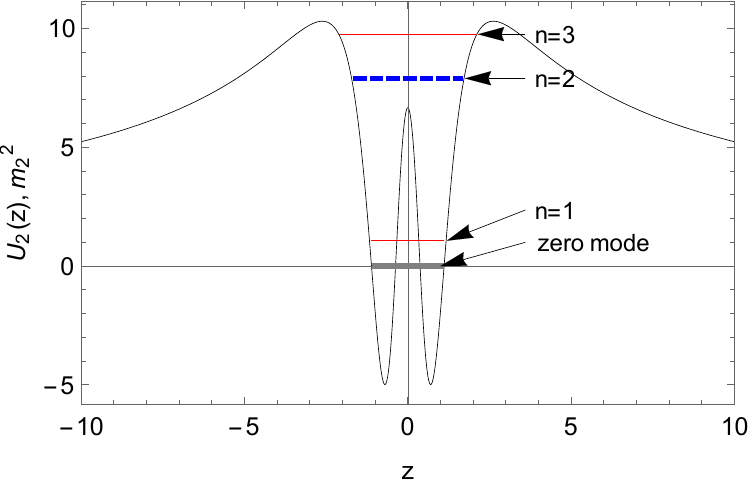}}
\hspace{0.5cm}
\subfigure[$t_1=20,t_2=0.25$.]{\label{2FigResPVolKR}
\includegraphics[width= 0.38\textwidth]{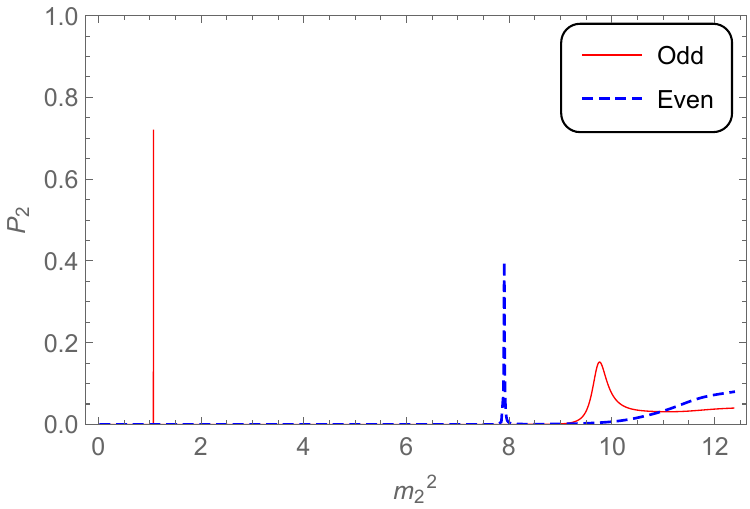}}
\subfigure[$t_1=30,t_2=0.25$.]{\label{3FigResSpecVolKR}
\includegraphics[width= 0.38\textwidth]{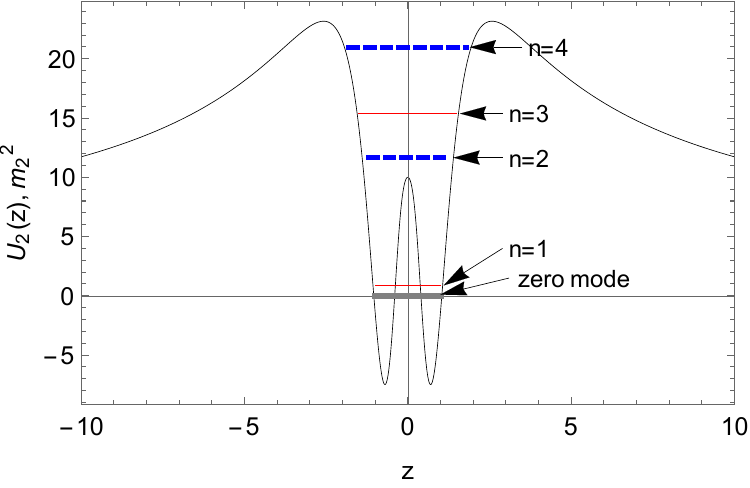}}
\hspace{0.5cm}
\subfigure[$t_1=30,t_2=0.25$.]{\label{3FigResPVolKR}
\includegraphics[width= 0.38\textwidth]{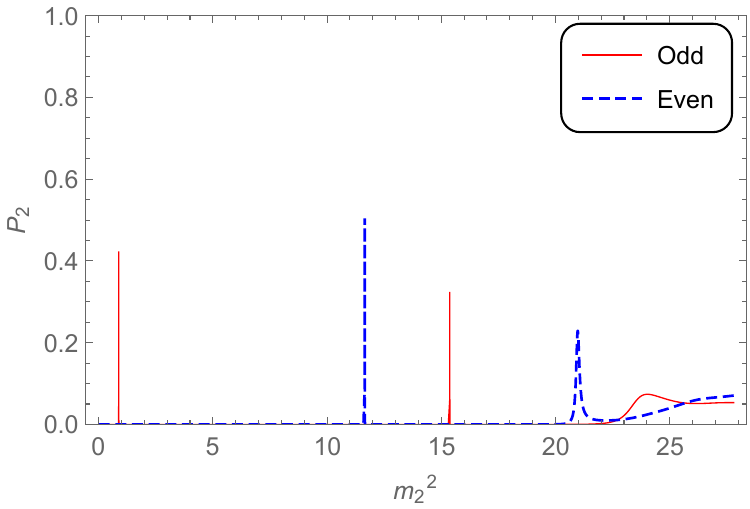}}
\end{center}\vskip -5mm
\caption{The mass spectra, the effective potential $U_2(z)$, and the corresponding relative probability
         $P_2$ with the parameters $k=1$, $v=3$ and $t_1=40,t_2=0.125$; $t_1=20,t_2=0.25$; $t_1=30,t_2=0.25$.
         The effective potential $U_2(z)$ for the black line, the zero mode for the grey line, the
         even-parity resonant KK modes for the blue lines, and the odd-parity resonant KK modes for the red
         lines.}
 \label{SpecPResoVolKR}
\end{figure}

\begin{figure} %[htb]
\begin{center}
\subfigure[$u_1$.]{\label{Odd1ResVolKR}
\includegraphics[width = 0.23\textwidth]{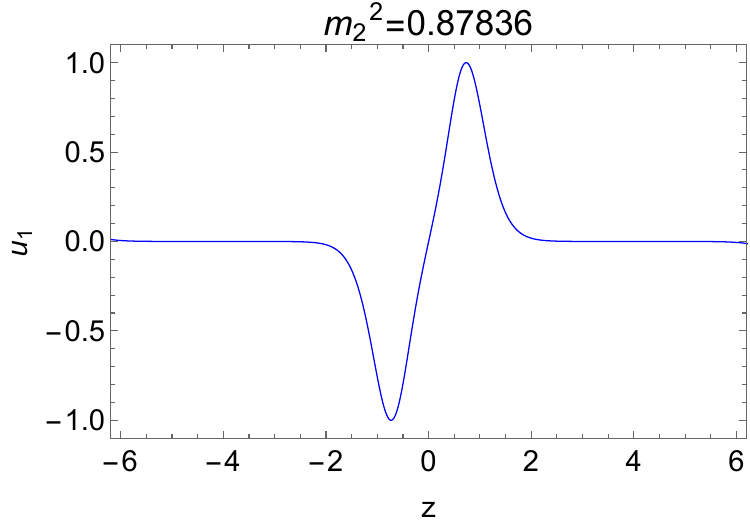}}
\subfigure[$u_1$.]{\label{Even1ResVolKR}
\includegraphics[width = 0.23\textwidth]{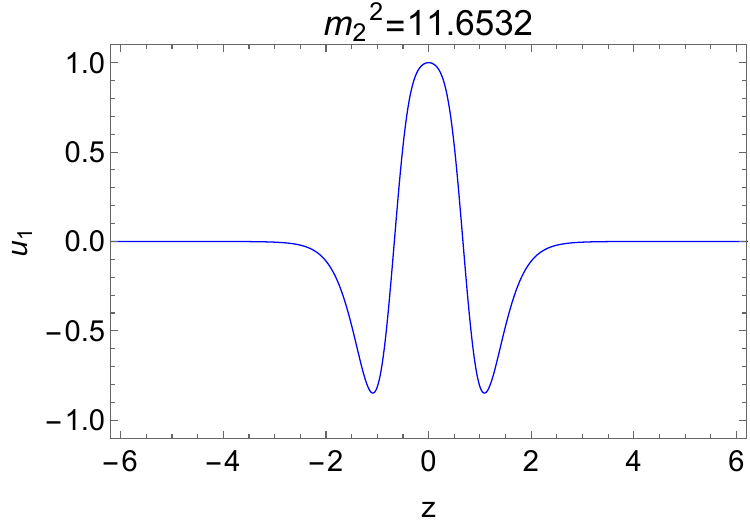}}
\subfigure[$u_1$.]{\label{Odd2ResVolKR}
\includegraphics[width = 0.23\textwidth]{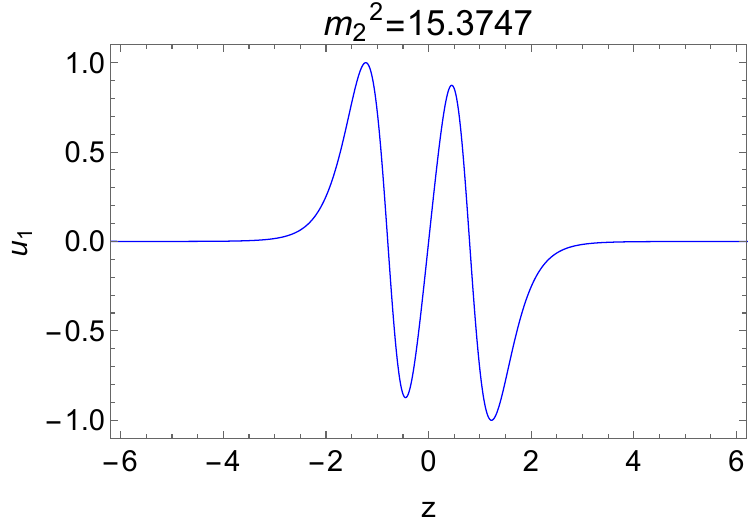}}
\subfigure[$u_1$.]{\label{Even2ResVolKR}
\includegraphics[width = 0.23\textwidth]{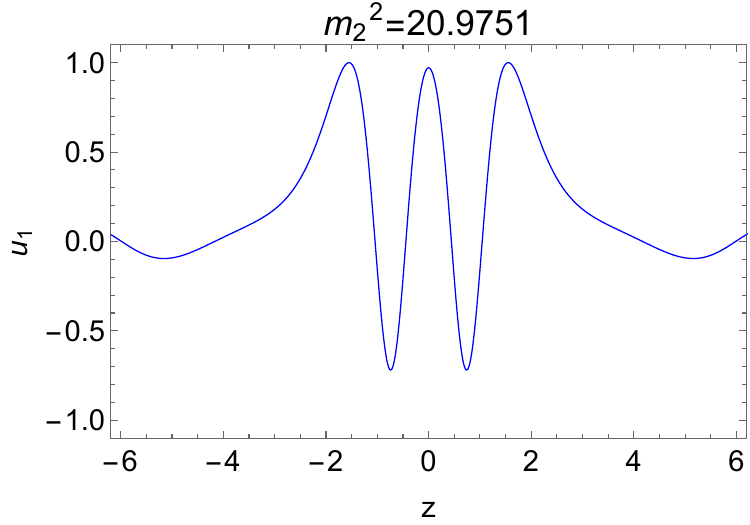}}
\end{center}\vskip -2mm
\caption{The shapes of resonant KK modes $u_1$ for the KR field with different $m_2^2$.
         The parameters are set as $k=1,v=3,t_1=30$ and $t_{2}=0.25$.}
 \label{FigResVolKR}
\end{figure}

First, consider the case $0<t_2<v^2/24$. Although none of the massive modes can be localized on
the brane, the effective potential develops a potential well around the brane position, allowing
massive modes to be quasi-localized as resonant modes. Using the relative probability method (\ref{PReso}),
we numerically determine the resonant modes of the effective potential (\ref{EffPotKR}). Their
mass spectra and the corresponding relative probabilities are displayed in fig. (\ref{SpecPResoVolKR}).
In the mass spectra, the zero mode is the ground state (bound state), and all massive modes
appear as resonant states. The masses, widths, and lifetimes of the obtained resonant modes
are listed in table \ref{tableRMOriKR}. The profiles of four representative resonant KK modes
for the parameter values $t_1=30$ and $t_2=0.25$ are shown in fig. \ref{FigResVolKR}. Therefore,
for $0<t_2<v^2/24$, the massive KK modes can be trapped near the brane as resonant states, and
the number of resonances increases with the coupling parameters $t_1$ and $t_2$.
%%%%%%%%%%%%%%%%%%%%%%%%%%%%%%%%%%%%%%%%%%%%%%%%%%%%%%%%%%%%%%%%%%%%%%%%%%%%
\begin{table}[tbp]
\centering
\begin{tabular}{|c|c|c|c|c|c|c|c|}
    \hline
    $t_1$               & $t_2$                 & $U^{\text{max}}_{2}$  & $n$         & $m_2^2$ &
    $m_2$               & $\Gamma$              & $\tau$
    \\
    \hline
    $40$                & $0.125$               & $8.0905$              & $1$         & $1.0589$   &
    $1.0290$            & $1.128\times10^{-7}$  & $8.863\times10^{6}$
    \\
                        &                       &                       & $2$         & $7.6332$ &
    $2.7628$            & $0.1075$              & $9.3062$
    \\
    \hline
    $20$                & $0.25$                & $10.3116$             & $1$         & $1.0678$ &
    $1.0334$            & $1.490\times10^{-9}$  & $6.711\times10^{8}$
    \\
    %\cline{3-9}
                        &                       &                       & $2$         & $7.9037$ &
    $2.8114$            & $2.965\times10^{-3}$  & $337.2954$
    \\
                        &                       &                       & $3$         & $9.7616$ &
    $3.1244$            & $0.0624$              & $16.0160$
    \\
    \hline
    $30$                & $0.25$                & $23.1669$             & $1$         & $0.8784$ &
    $0.9372$            &$2.862\times10^{-10}$  & $3.494\times10^9$
    \\
                        &                       &                       & $2$         & $11.6532$ &
    $3.4137$            & $4.441\times10^{-7}$  & $2.252\times10^6$
    \\
                        &                       &                       & $3$         & $15.3747$ &
    $3.9211$            & $2.555\times10^{-5}$  & $3.914\times10^4$
    \\
                        &                       &                       & $4$         & $20.9751$ &
    $4.5799$            & $0.0195$              & $51.2958$
    \\
    \hline
\end{tabular}
\caption{The masses, widths, and lifetimes of resonant KK modes $u_1(z)$.
         The parameters are set as $k=1$ and $v=3$. }
    \label{tableRMOriKR}
\end{table}
%%%%%%%%%%%%%%%%%%%%%%%%%%%%%%%%%%%%%%%%%%%%%%%%%%%%%%%%%%%%%%%%%%%%%%%%%%%%%%%%%%%%%

\begin{figure} %[htbp]
\begin{center}
\includegraphics[width= 0.49\textwidth]{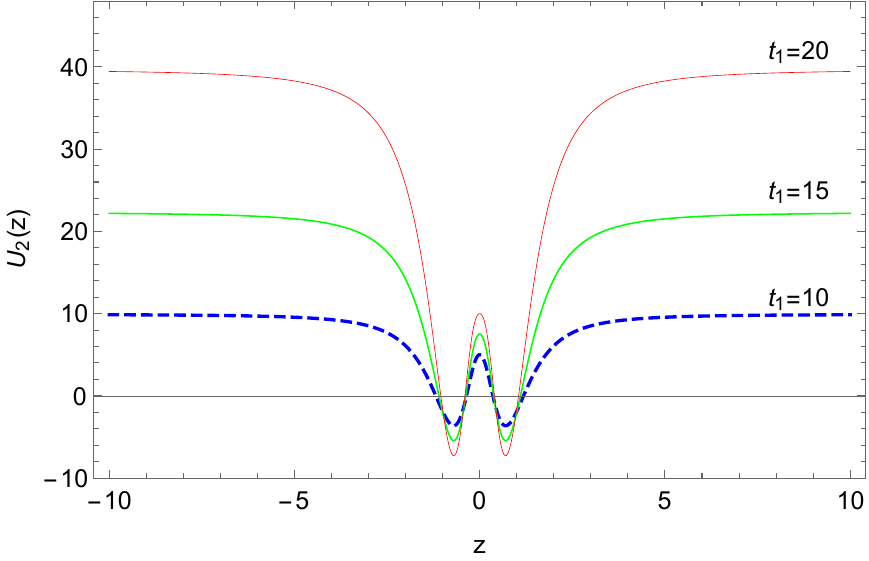}
\caption{The effective potential $U_2(z)$ with the parameter $t_1=10,15$, and $20$.
        The other parameters are set as $k=1,v=3$, and $t_2=3/8$.}
\label{figSpecPTKR}
\end{center}
\end{figure}

Next, consider the case $t_2=v^2/24$, for which the effective potential takes the form of a PT potential.
In this scenario, a finite number of massive KK modes can be localized on the brane. The asymptotic
value of the effective potential, given by eq. (\ref{LmtPTKR}), depends on the coupling parameter $t_1$, which
determines the number of localized massive modes. The effective potential is depicted numerically
in fig. \ref{figSpecPTKR} for the parameter values $k=1,v=3,t_2=3/8$ and $t_1=10,15,20$. As shown
in the figure, the potential well becomes deeper as $t_1$ increases. For each potential profile,
the corresponding mass spectra are solved numerically, as listed follows
\begin{eqnarray}
  m_2^2=&\hspace{-4.3cm}\{0,1.11,6.05,7.53,9.18,9.63,9.88\},                                      &\text{for}~t_1=10;     \label{SpecPTKR1} \\
  m_2^2=&\hspace{-0.1cm}\{0,1.04,8.99,11.89,16.16,18.52,20.42,21.38,21.94,22.17\},\hspace{0.3cm}  &\text{for}~t_1=15;     \label{SpecPTKR2} \\
  m_2^2=&\hspace{-0.4cm}\{0,0.89,11.61,15.53,22.05,26.71,30.87,33.95,36.18,37.66,                 &\nonumber                                \\
        &\hspace{-5.5cm}38.59,39.12,39.41,39.57\},                                                &\text{for}~t_1=20.     \label{SpecPTKR3}
\end{eqnarray}
These results indicate that the number of localized massive KK modes increases with the coupling
parameter $t_1$. Therefore, in the case $t_2=v^2/24$, both the zero mode and a finite number of
massive KK modes can be localized on the brane.

\begin{figure} %[htbp]
\begin{center}
\includegraphics[width= 0.49\textwidth]{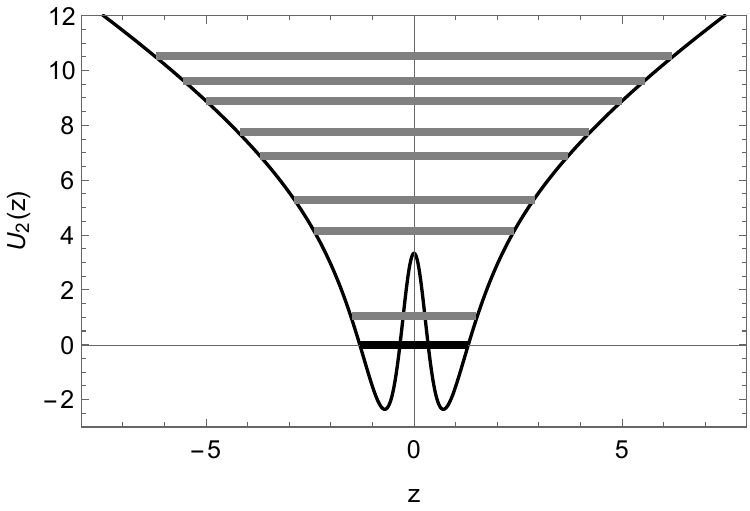}
\caption{The effective potential $U_2(z)$ with the parameters $k=1,v=3,t_1=5$, and $t_2=0.5$.}
\label{figSpecIDWKR}
\end{center}
\end{figure}

When the coupling parameter is chosen as $t_2=0.5>v^2/24$, the effective potential becomes an
infinitely deep potential well. Consequently, all massive modes are bound states. The mass spectrum
of the low-lying localized KK modes is displayed in fig. \ref{figSpecIDWKR} for $k=1,v=3,t_1=5$ and
$t_2=0.5$. As expected for an infinitely deep potential well, all massive modes are localized on the
brane, leading to an infinitely discrete spectrum of mass.

Therefore, by introducing the coupling between the 6D KR field and the gravity, the zero mode of
the KR field can be localized on the thick flat brane. The behavior of the massive KK modes depends on the
value of the coupling parameter $t_2$: they can either be quasi-localized as resonant states or
fully localized on the brane. For $0<t_2<v^2/24$, the effective potential exhibits a volcano-like
shape, and no massive KK modes can be localized on the brane. However, some massive modes may
appear as resonant states and thus remain quasi-localized near the brane. When $t_2=v^2/24$, the
effective potential approaches a positive constant when far away from the brane. In this case, a
finite number of massive KK modes can be localized on the brane, and the number of localized massive
modes increases with the coupling parameter $t_1$. For $t_2>v^2/24$, the effective potential becomes
an infinitely deep potential well. All massive modes are localized on the thick brane.

%%%%%%%%%%%%%%%%%%%%%%%%%%%%%%%%%%%%%%%%%
\section{Conclusions}\label{Cons}

In this paper, we investigate the localization of $q$-form fields on a codimension-two thick flat brane. To
account for the interaction between the bulk fields and the gravity, we introduce a nonminimal
coupling between the 6D $q$-form fields and the gravity through a curvature-dependent coupling
function $F(R)$. The function $F(R)$ depends on the bulk scalar curvature $R$ and contains two
positive coupling parameters $t_1$ and $t_2$, which describe the strength of the coupling. Based
on this setup, we analyze the localization and KK spectra of different $q$-form fields, namely the
scalar field, the $U(1)$ gauge vector field and the KR field.

The 6D spacetime considered in this work consists of a 3-brane and two extra dimensions, one of
which is noncompact (large) and the other compact. Since the 6D brane configuration is regular
and free of singularities, the KK modes of various $q$-form fields are automatically normalizable
along the compact extra dimension. Consequently, the localization problem reduces to examining the
localization behavior of these KK modes along the large extra dimension.

Starting from the action of the 6D $q$-form fields, we derive the Schr\"{o}diner-like equations
governing their KK modes by applying the variational principle. The corresponding zero mode solutions
can then be obtained from these equations. An important observations is that different $q$-form fields
share the same qualitative localization behavior. Specifically, for positive coupling parameters $t_1$
and $t_2$, the zero modes of all $q$-form fields are normalizable and can therefore be localized on
the thick flat brane.

Furthermore, the effective potentials associated with the massive KK modes of these $q$-form fields
manifest three distinct asymptotic behaviors, depending on the value of the coupling parameter $t_2$.
For $0<t_2<v^2/24$, the effective potentials take the form of volcano-like potentials. In this case,
the massive KK modes of different $q$-form fields cannot be localized on the brane. Nevertheless,
they may exist as resonant KK modes, and the number of such resonances increases with the coupling
parameters $t_1$ and $t_2$. For $t_2=v^2/24$, the effective potentials reduce to PT potentials and
approach a positive constant when far away from the brane. Consequently, a finite number of massive
KK modes can be localized on the thick brane. For $t_2>v^2/24$, the effective potentials belong to
the infinitely deep potential wells. As a result, all massive KK modes of the $q$-form fields are
localized on the thick brane, giving rise to infinitely discrete spectra of mass. In addition, no
tachyonic modes are found in the KK spectra of these $q$-form fields, indicating the stability of
the corresponding localized configurations.

\acknowledgments

This work is supported by the National Natural Science Foundation of China (Grants No. 11305119),
the Natural Science Basic Research Plan in Shaanxi Province of China (Program No. 2020JM-198),
and the 111 Project (B17035).

\end{document}